\documentclass{aa}

\usepackage{graphicx}
\defcitealias{Hammond2022}{Hammond et al.}

\usepackage{txfonts}
\usepackage[colorlinks=true,linkcolor=blue,citecolor=blue]{hyperref}
\begin{document}

   \title{A Mini-Chemical Scheme with Net Reactions for 3D GCMs}

   \subtitle{II. 3D thermochemical modelling of WASP-39b and HD 189733b}

   \author{Elspeth K.H. Lee\inst{1}, 
          Shang-Min Tsai\inst{2,3},
        Mark Hammond\inst{3}
        \and Xianyu Tan\inst{4,5,3}
          }

   \institute{Center for Space and Habitability, University of Bern, Gesellschaftsstrasse 6, CH-3012 Bern, Switzerland \\
              \email{elspeth.lee@unibe.ch}
         \and
            Department of Earth Sciences, University of California, Riverside, California, US \\
            \email{shangmit@ucr.edu}
         \and
             Atmospheric, Ocean, and Planetary Physics, Department of Physics, University of Oxford, UK
        \and
          Tsung-Dao Lee Institute, Shanghai Jiao Tong University, 520 Shengrong Road, Shanghai, People’s Republic of China
        \and 
          School of Physics and Astronomy, Shanghai Jiao Tong University, 800 Dongchuan Road, Shanghai, People’s Republic of China
             }

   \date{Received 16 November 2022 / Accepted 17 February 2023}

\titlerunning{3D GCM mini chem}


  \abstract
  {The chemical inventory of hot Jupiter (HJ) exoplanets atmospheres continue to be observed by various ground and space based instruments in increasing detail and precision.
  It is expected that some HJs will exhibit strong non-equilibrium chemistry characteristics in their atmospheres, which might be inferred from spectral observations.}
   {We aim to model the three dimensional thermochemical non-equilibrium chemistry in the atmospheres of the HJs WASP-39b and HD 189733b.}
   {We couple a lightweight, reduced chemical network `mini-chem' that utilises net reaction rate tables to the Exo-FMS General Circulation Model (GCM).
   We perform GCM models of the exoplanets WASP-39b and HD 189733b as case studies of the coupled mini-chem scheme.
   The GCM results are then post-processed using the 3D radiative-transfer model gCMCRT to produce transmission and emission spectra to assess the impact of non-equilibrium chemistry on their observable properties.}
   {Both simulations show significant departures from chemical equilibrium (CE) due to the dynamical motions of the atmosphere.
   The spacial distribution of species generally follows closely the dynamical features of the atmosphere rather than the temperature field.
   Each molecular species exhibits a different quench level in the simulations, also dependent on the latitude of the planet.
   Major differences are seen in the transmission and emission spectral features between the CE and kinetic models.}
   {Our simulations indicate that considering the 3D kinetic chemical structures of HJ atmospheres has an important impact on physical interpretation of observational data. Drawing bulk atmospheric parameters from fitting feature strengths may lead to inaccurate interpretation of chemical conditions in the atmosphere of HJs. Our open source mini-chem module is simple to couple with contemporary HJ GCM models without substantially increasing required computational resources.}


   \maketitle
%
\section{Introduction}
\label{sec:intro}

The chemical complexity of hot Jupiter exoplanets continues to be revealed in ever increasing detail and accuracy by observational efforts.
Observational data and theoretical studies in recent years have shown the importance of considering the 3D nature of exoplanet atmospheres when making interpretations of their physical mechanisms at play \citep[e.g.][]{Knutson2007}.
Of this, the 3D distribution and effects of dynamics on the chemical structures of exoplanet atmospheres has become an important consideration when interpreting different sources of observational data for a single planet or inferring chemical trends between planets \citep[e.g.][]{Steinrueck2019,Drummond2020}.

Ground based high spectral resolution studies have detected a multitude of chemical species across the hot Jupiter population \citep[e.g.][]{Ehrenreich2020,Merritt2020,Hoeijmakers2020,Line2021,Giacobbe2021,Prinoth2022}, showing the massive variety in the chemical inventory in these atmospheres and differences between the dayside and east/west terminator region compositions.
The next generation of high resolution instruments on board thirty meter class telescopes (e.g. EE-ELT: ANDES; \citet{Marconi2022}) is likely to greatly improve the signal to noise of such observations.

JWST is slated to drastically improve the accuracy and wavelength coverage when characterising exoplanet atmospheres compared to capabilities of the Hubble Space Telescope (HST).
JWST has observed three hot gas giant planets (WASP-18b, WASP-39b and WASP-43b) as part of the Early Release Science (ERS) program \citep{Bean2018} across its broad wavelength range capabilities in an effort to assess in detail the chemical inventory of these atmospheres.
In conjunction during JWST's lifetime, the near-future ARIEL mission \citep{Tinetti2016, Charnay2021, Tinetti2021} will characterise 100s to 1000s of exoplanet atmospheres during its operational lifetime and enable statistical studies of the exoplanet population.

The modelling of chemical species in hot Jupiter atmospheres has various levels of complexity, from assuming all species are in chemical equilibrium (CE) to large chemical networks where the conversion between species is directly evolved in time simultaneously.
To date, most General Circulation Models (GCMs) have assumed some form of CE in their modelling efforts, typically assumed when applying real gas radiative-transfer (RT) modelling.
Either assuming CE during the simulation runtime to calculate gas abundances for RT schemes \citep[e.g.][]{Amundsen2016, Drummond2018} or during the construction of `pre-mixed' opacity tables \citep[e.g.][]{Showman2009, Lee2021, Schneider2022, Deitrick2022}.
Post-processing efforts of GCM results have also used CE abundances extensively in the literature \citep[e.g.][]{Fortney2010, Lines2018, Lee2021, Schneider2022, Robbins-Blanch2022}.
Models also typically include local condensation of species through local depletion of elements, or ad-hoc removal of species (usually TiO and VO and other strong UV and optical absorbers) that may have `rained out' from photospheric regions. 
CE assumes that the chemical inventory of a volume is time-independent, and that the chemical reactions have had infinite time to relax to their equilibrium state.
Many groups have developed CE schemes which are used often by members of the community \citep[e.g.][]{Visscher2010, Blecic2016, Woitke2018, Stock2022} or make use of the NASA CEA code \citep{McBride1996}.

Chemical kinetic models investigating the chemical composition of hot Jupiter exoplanets have mostly been performed in 1D \citep[e.g.][]{Moses2011, Venot2012, Rimmer2016, Tsai2017}, which have enlightened the field greatly to the important role vertical mixing and photochemistry in the atmosphere play in setting the chemical compositions.
These models tend to contain the largest and most complete chemical networks in their methodology.
Recently, pseudo-2D models utilising the output of GCM models have been used to detail the composition of hot Jupiter planets at the equatorial regions \citep[e.g.][]{Agundez2012, Agundez2014, Venot2020,Baeyens2021, Moses2021, Baeyens2022}.
These studies have shown the importance of horizontal quenching in addition to vertical quenching, typically homogenising the mixing ratio of species across longitude from the dayside chemical composition.

The chemical relaxation method has been widely used to simplify the modelling of complex chemical networks.
Many groups have investigated chemical non-equilibrium effects using the chemical relaxation method, first described for hot Jupiters in \citet{Cooper2006}.
The chemical relaxation scheme evolves the formula
\begin{equation}
\frac{D X_{i}}{D t} = \frac{X_{i} - X_{i, eq}}{\tau_{\rm chem}},
\end{equation}
where X$_{i}$ is the mixing ratio of the species i, X$_{i, eq}$ the mixing ratio at chemical equilibrium and $\tau_{\rm chem}$ [s] the chemical timescale parameter.
This effectively relaxes individual species towards their local chemical equilibrium values according the species chemical timescale.
This chemical timescale can be estimated from the rate limiting step of a specific species, by examination of a full kinetic network which is temperature and pressure dependent \citep{Tsai2018}.
Example studies that have investigated this method with various sophistication include \citet{Drummond2018b, Tsai2018, Mendonca2018} and \citet{Steinrueck2019}.

The kinetic network chemistry method directly integrates the reaction rates between different species simultaneously together in a `chemical network'.
The local evolution of the number density of species $i$, n$_{i}$ [cm$^{-3}$], is given by \citep[e.g.][]{Rimmer2016,Tsai2018}
\begin{equation}
\label{eq:kin_chem_eq}
\frac{D n_{i}}{D t} = P_{i} - L_{i} - \nabla \phi_{i},
\end{equation}
where P$_{i}$ [cm$^{-3}$ s$^{-1}$] are the reactions that source (produce) species i (i.e. increasing it's abundance), and L$_{i}$ [cm$^{-3}$ s$^{-1}$] are reactions that reduce (lose) species i (i.e. decreasing it's abundance).
$\phi_{i}$ [cm$^{-3}$ s$^{-1}$] represents the net flux of species $i$ in and out of the volume from dynamical motions.
The production or loss of a species depends on the local chemical and thermodynamic properties of the considered region of atmosphere.
Since the advective flux of species is calculated by the advection of tracers within the GCM dynamical core which resolves large scale transport and overturning, we do not model the mixing of species through an eddy diffusive process, typically represented by a diffusive term in Eq. \eqref{eq:kin_chem_eq}, characterised by an atmospheric diffusion coefficient K$_{\rm zz}$ \citep[e.g.][]{Rimmer2016} or a diffusive and/or turnover mixing timescale $\tau_{\rm mix}$ \citep[e.g.][]{Woitke2020} common to 1D modelling efforts.

Coupling kinetic models directly to 3D HJ GCM models has been proven to be a challenge and has so far only been attempted by \citet{Drummond2020} and \citet{Zamyatina2023}.
Both studies coupled the UM GCM \citep{Mayne2014} to a network of 30 species with 181 reactions published by \citet{Venot2019}, reduced from the full model of 105 species with $\approx$1000 reactions from the \citet{Venot2012} network.
They also included radiative-feedback from the changing mixing ratios of gas species due to the disequilibrium brought on by atmospheric dynamics.
In addition to vertical and zonal quenching, their study also showed the importance of meridional quenching effects, suggesting that the 3D dynamical structures present in the atmosphere are a vital consideration in setting the global chemical inventory and spacial distribution.
\citet{Zamyatina2023} were able to expand the \citet{Drummond2020} study by simulating HAT-P-11b and WASP-17b in addition to the canonical HD 189733b and HD 209458b.
However, such GCMs with comprehensive chemical network are too computationally expensive to conduct a wider range of numerical survey which is key to improve our understanding of 3D chemical transport. 
Therefore, developing a GCM framework with a more portable chemical scheme such as the \citet{Tsai2022} net reaction table scheme is urgently needed.

In this follow up paper to \citet{Tsai2022}, we couple the mini-chem chemical network to the 3D Exo-FMS GCM in the hot Jupiter setup \citep{Lee2021}.
We then use the hot Jupiters WASP-39b and HD 189733b as testbeds of the network, examining the non-equilibrium chemistry behaviour of both simulations.

WASP-39b is a low gravity hot Saturn exoplanet, discovered by \citet{Faedi2011}.
It has been characterised in transmission by \citet{Wakeford2018} who combined new HST WFC3 data with previous HST STIS, Spitzer and VLT FORS2 data.
Recently, it was part of the DD-ERS JWST program \citep{Batalha_JWST,Bean2018}, with transmission spectroscopy from the NIRSpec PRISM \citep{JWST2022,Rustamkulov2022} and G395H modes \citep{Alderson2022}, as well as NIRISS SOSS \citep{Feinstein2022} and NIRCam F322W2 \citep{Ahrer2022}.
A major discovery of this group effort was the detection of SO$_{2}$, a photochemically produced species, suggesting that non-equilibrium chemical processes are an important factor to consider when characterising exoplanet atmospheres.
Modelling of the photochemical environment with a focus on SO$_{2}$ in WASP-39b in the context of the JWST observational data was performed in \citet{Tsai2023}.

HD 189733b is a well known canonical HJ discovered by \citet{Bouchy2005}. It is scheduled to be visited by several cycle 1 JWST programs \citep{Deming_JWST,Min_JWST,Kilpatrick_JWST}.
It is one of the most characterised HJ exoplanets to date, with extensive HST \citep[e.g.][]{Pont2013,McCullough2014, Sing2016} transmission and Spitzer \citep[e.g.][]{Charbonneau2008, Knutson2009} emission and phase curve data available.

After each simulation is completed, we post-process the model outputs using the 3D radiative-transfer model gCMCRT \citep{Lee2022b} to produce transmission, emission and phase curve spectra to examine the overall affect of non-equilibrium chemistry on the observational properties of each atmosphere.
In section \ref{sec:mini_chem}, we introduce the mini-chem methodology and implementation into the Exo-FMS GCM.
In section \ref{sec:GCM_modelling} we present the GCM modelling aspects of our study for the WASP-39b and HD 189733b.
In section \ref{sec:results} we show the GCM results of our study for WASP-39b and HD 189733b.
In section \ref{sec:pp} we show the transmission, emission and phase curve post-processing efforts using the results of the coupled GCM.
Section \ref{sec:drummond} compares our current results to those found in \citet{Drummond2020}.
Section \ref{sec:discussion} presents a discussion of our GCM and post-processing results.
Section \ref{sec:conclusion} shows a summary and conclusions of our study.

\section{Mini-chem methodology}
\label{sec:mini_chem}

\begin{table}
\centering
\caption{List of reactions and rate coefficients used in the NCHO net chemical network. `M' denotes a neutral third body species.
See \citet{Tsai2022} for the rate coefficients.}
\begin{tabular}{c c c c }  \hline \hline
Forward Reaction & T$_{\rm min}$ & T$_{\rm max}$ &  Notes \\ \hline
OH + H$_{2}$ $\rightarrow$ H$_{2}$O +  H & 250 & 2580 & Elementary   \\
OH + CO $\rightarrow$ CO$_{2}$ + H & 300 & 2000 & Elementary  \\
O + H$_{2}$ $\rightarrow$ OH + H & 300 & 2500 &  Elementary  \\
H + H + M $\rightarrow$ H$_{2}$ + M & 100 & 5000 &  Elementary  \\
CH$_{4}$ + H$_{2}$O  $\rightarrow$ CO + 3H$_{2}$ & 300 & 3451 &  Net \\
CH$_{4}$ + CH$_{4}$  $\rightarrow$ C$_{2}$H$_{2}$ + 3H$_{2}$ & 300 & 3451 & Net \\
CH$_{4}$ + CO  $\rightarrow$ C$_{2}$H$_{2}$ + H$_{2}$O & 300 & 3451  & Net \\
NH$_{3}$ + NH$_{3}$  $\rightarrow$ N$_{2}$ + 3H$_{2}$ & 300 & 3451 & Net \\
CH$_{4}$ + NH$_{3}$  $\rightarrow$ HCN + 3H$_{2}$ & 300 & 3451 &  Net \\
CO + NH$_{3}$  $\rightarrow$ HCN + H$_{2}$O & 300 & 3451 &  Net \\ \hline
\end{tabular}
\label{tab:networks}
\end{table}

`Mini-chem' is a publicly available chemical kinetics solver package, designed to be lightweight and simple to couple to contemporary GCM models, or other types of 3D dynamical models used by the community.
The standalone 0D version, written in Fortran 90, is available from the lead author's GitHub \footnote{\url{https://github.com/ELeeAstro}} which can be readily adapted to suit each GCM modelling needs.
We have also made some stiff ordinary differential equation (ODE) solvers `threadsafe', able to be run with OpenMP to increase the efficiency inside GCMs that use the OpenMP framework.
Mini-chem currently has three miniature chemical networks available, HO, CHO and NCHO, with the CHO and NCHO networks making use of the net reaction rate tables tested and benchmarked in \citet{Tsai2022}.
A table of the reactions and reference used in the mini-chem networks are provided in Table \ref{tab:networks}.
In this study, we focus on the NCHO model and its coupling to the GCM models.

Mini-chem is separated into three main parts:
\begin{itemize}
\item The input routine that reads species data, rate coefficients and net-rate tables.
\item The chemistry routines that finds the forward reaction rates and the reverse rates using the thermochemical data of the involved species \citep{Tsai2017}, for a given atmospheric pressure and temperature.
\item The stiff ODE solver, that integrates the chemical species together in time every chemical timestep.
\end{itemize}

What makes mini-chem different from full kinetic chemistry models is the use of `net reaction rate' tables, developed as offshoots from the full VULCAN photochemical kinetic model \citep{Tsai2017, Tsai2018, Tsai2022}.
These tables take the form of temperature and pressure dependent net forward reaction rates.
For example, tables of the net reaction rate CH$_{4}$ + H$_{2}$O $\rightarrow$ CO + 3H$_{2}$ are created, which are then interpolated to the atmospheric temperature and pressure inside the GCM to find the forward reaction rate of this net reaction.
Since the tables are temperature and pressure dependent, the forward rate coefficient changes dependent on the most efficient chemical pathway calculated through the full chemical kinetic network.
One caveat of this scheme is that a metallicity must be assumed to generate the tables, making the tables only valid for a single metallicity value.
Our tables span a pressure range of 10$^{-8}$-1000 bar and temperature range of 300-3451 K. The net tables are then interpolated to the local gas temperature and pressure using 2D Bezier interpolation \citep[e.g.][]{Hennicker2020} during runtime.
As such, in this study we assume 10x solar metallicity for WASP-39b and solar metallicity for HD 189733b.
The temperature and pressure dependent reverse reaction rates are then calculated using the equilibrium constants of the reaction, where we use the NASA polynomial tables\footnote{\url{http://garfield.chem.elte.hu/Burcat/burcat.html}} to derive the equilibrium constants.
In App. \ref{sec:mc_contours} we present contour plots of the net forward reaction rate tables for solar metallicity.

Mini-chem therefore retains the architecture of a traditional kinetic chemistry scheme, using a stiff ODE solver to evolve the number density of each species simultaneously in time during the simulation, but with a much reduced species and reaction list compared to a full network.
For the stiff ODE solver inside the Exo-FMS GCM, we use the \textsc{seulex}\footnote{\url{http://www.unige.ch/~hairer/software.html}} implicit Euler solver developed by \citet{Hairer2010}.
However, the standalone GitHub version includes many ODE solver variants for testing purposes and may be more suitable for coupling to other GCM models or for additional studies.
Calculations are made more efficient and accurate in mini-chem by including the Jacobian matrix of each network into the ODE solvers.
 
Our mini-chem method \citep{Tsai2022} reduces the number of species and GCM tracers to 12 (OH, H$_{2}$, H$_{2}$O, H, CO, CO$_{2}$, O, CH$_{4}$, C$_{2}$H$_{2}$, NH$_{3}$, N$_{2}$ and HCN) with 10 reactions (Tab. \ref{tab:networks}) for the NCHO model.
This makes mini-chem a practical and computationally light model compared to the \citet{Drummond2020} study which used 30 species with 181 reactions for their scheme.
The mixing ratio of each species is advected quantity within the dynamical core.
Our model is able to be included inside contemporary gas giant GCM models in a simple manner without too much additional computational effort.
This method offers useful middle ground between the expensive full kinetic schemes which require 100s of species and chemical tracers to be evolved in the GCM and the chemical relaxation method, where chemical timescale parameters must be estimated which are generally more inaccurate.

We report approximately 4-5 minutes of walltime using 58 processors to simulate 1 Earth day\footnote{Using a server with Intel Xeon E5-2698 v3 2.30GHz and Intel Xeon Gold 6130 CPU 2.10GHz CPUs}.
This is about twice as fast than \citet{Drummond2020} who report about one week of walltime using $\approx$200 processors to simulate 1000 Earth days\footnote{Using the DiRAC DIaL supercomputing facility} (approximately 10 minutes per Earth day).
Differences in the RT scheme, GCM spatial resolution, innate GCM performance, implementation and compute server differences between the studies make an accurate comparison of runtimes difficult.
However, assuming we used the same chemical timestep as \citet{Drummond2020} and that the picket fence scheme running every timestep is equivalent to their corr-k model every 5 timesteps, our scheme is approximating 4 to 6 times faster than \citet{Drummond2020}.
\citet{Tsai2022} report an approximate 10 times speedup using the mini-chem network compared to a similar sized network to \citet{Venot2019} on the same computing hardware.
Overall, our experience coupling the scheme and results here show that mini-chem is suitable for running on modest computational resources and does not require a large leap in resources beyond the original GCM model requirements to run.

\section{GCM modelling}
\label{sec:GCM_modelling}

\begin{table*}
\centering
\caption{Adopted GCM simulation parameters for WASP-39b and HD 189733b.}
\begin{tabular}{c c c c l}  \hline \hline
 Symbol & WASP-39b & HD 189733b & Unit & Description \\ \hline
 T$_{\rm int}$ & 358 & 382 & K & Internal temperature \\
 T$_{\rm irr}$ & 1652 & 1694 & K & Irradiation temperature \\
 P$_{\rm 0}$ & 220 & 220 &  bar & Reference surface pressure \\
 M/H & 10x & 1x & -  & Metallicity times solar \\
 c$_{\rm P}$ & 11335 & 12637 &  J K$^{-1}$ kg$^{-1}$ & Specific heat capacity \\
 R & 3221 & 3561 &  J K$^{-1}$ kg$^{-1}$  & Ideal gas constant \\
 $\kappa$ & 0.284 & 0.282 & -  & Adiabatic coefficient \\
 g$_{\rm p}$ & 4.26 & 21.8 & m s$^{-2}$ & Acceleration from gravity \\
 R$_{\rm p}$ & 9.14 $\times$ 10$^{7}$ & 8.14 $\times$ 10$^{7}$ & m & Radius of planet \\
 $\Omega_{\rm p}$ & 1.79 $\times$ 10$^{-5}$ &  3.38 $\times$ 10$^{-5}$& rad s$^{-1}$ & Rotation rate of planet \\
 $\Delta$t$_{\rm hydro}$ & 30 & 30 & s & Hydrodynamic time-step \\
 $\Delta$t$_{\rm rad}$  & 30 & 30 & s & Radiative time-step \\
 $\Delta$t$_{\rm chem}$  & 1500 & 1500 & s & Chemical time-step \\
 N$_{\rm v}$ & 54 & 54 & - & Vertical resolution \\
 d$_{\rm 4}$ & 0.16 & 0.16 & - & $\mathcal{O}$(4) divergence dampening coefficient \\
\hline
\end{tabular}
\label{tab:GCM_parameters}
\end{table*}

In this study, we couple mini-chem to the Exo-FMS GCM model, which has been used to model atmospheres from hot rocky lava planets \citep{Hammond2017} to highly irradiated brown dwarf atmospheres \citep{Lee2020}.
Exo-FMS has been benchmarked to other GCM studies for the hot Jupiter regimes, showing it reproduces the gross atmospheric properties from other GCM modelling groups \citep{Lee2021}.
Following the advice in \citetalias{Hammond2022} \citeyear{Hammond2022}, we switch to a 4th order divergence dampening scheme to maintain numerical stability, rather than the 2nd order used in the previous Exo-FMS hot Jupiter studies.
This makes the dampening more scale selective, avoiding larger global scale dampening in favour of dampening smaller scale flows.
We do not include any explicit diffusion in our GCM, but include a Rayleigh basal drag of 1 Earth day \citep[e.g.][]{Carone2020} at the high pressure boundary at 220 bar in the simulations, decreasing linearly to 100 bar.
This ad-hoc localised drag is added to aid numerical stability and smoother transport of tracers near the simulation lower boundary.
More realistic drag profiles in GCMs, such as that produced from magnetic field sources, has been investigated in several studies \citep[e.g.][]{Perna2010, Beltz2022}.

Our RT package inside the GCM for this study consists of using the `adding method' \citep[e.g.][]{Mendonca2015} for the shortwave radiation, which includes the effects of scattering.
For the longwave radiation we use the short characteristics method \citep{Olson1987} with Bezier interpolants \citep[e.g.][]{delaCruz2013, Hennicker2020} instead of the linear interpolants for the source function estimate.
This is a fast, stable and highly accurate two-stream method but our current implementation does not include scattering for longwave radiation.
However, longwave radiation is most directly scattered by the presence of aerosol material which is not considered in this study.
We also use Bezier interpolation \citep[e.g.][]{Hennicker2020} to interpolate the temperature at the layers, as given by the GCM model, to the temperature at the levels to facilitate the longwave two-stream calculations.
We include the correction of pseudo-spherical geometry on the incident, height dependent, zenith angle following \citet{Li2006} and \citet{Mendonca2018b}.

Throughout this study, we use the picket-fence gas opacity scheme \citep{Chandrasekhar1960} developed for gas giant atmospheres by \citet{Parmentier2014} and  \citet{Parmentier2015}.
This scheme is efficient enough to be called every dynamical timestep and reproduces well the temperature and dynamical structures of the atmosphere when compared to a full correlated-k model.
We assume a hot internal temperature of T$_{\rm int}$ = 358 and 382 K for WASP-39b and HD 189733b respectively.
These values were taken from the \citet{Thorngren2019} relationship between irradiated flux and internal temperature.
More details on this implementation for Exo-FMS can be found in \citet{Lee2021}.
The standalone radiative-transfer routines used in Exo-FMS are publicly available from the lead author's GitHub \footnote{\url{https://github.com/ELeeAstro}}.
Dry convective adjustment is included as described in \citet{Lee2021}. 

For the initial chemical abundances for each chemical tracer species inside the GCM, we interpolate in T-p from a CE table computed from the GGchem \citep{Woitke2018} chemical equilibrium solver.
The GCM parameters used for our WASP-39b and HD 189733b simulations can be found in Table \ref{tab:GCM_parameters}.

\section{3D thermochemical kinetic compositions}
\label{sec:results}

In this section, we present the results of the WASP-39b and HD 189733b GCM simulations coupled to the mini-chem scheme. 

\subsection{WASP-39b}

\begin{figure*} 
   \centering
   \includegraphics[width=0.49\textwidth]{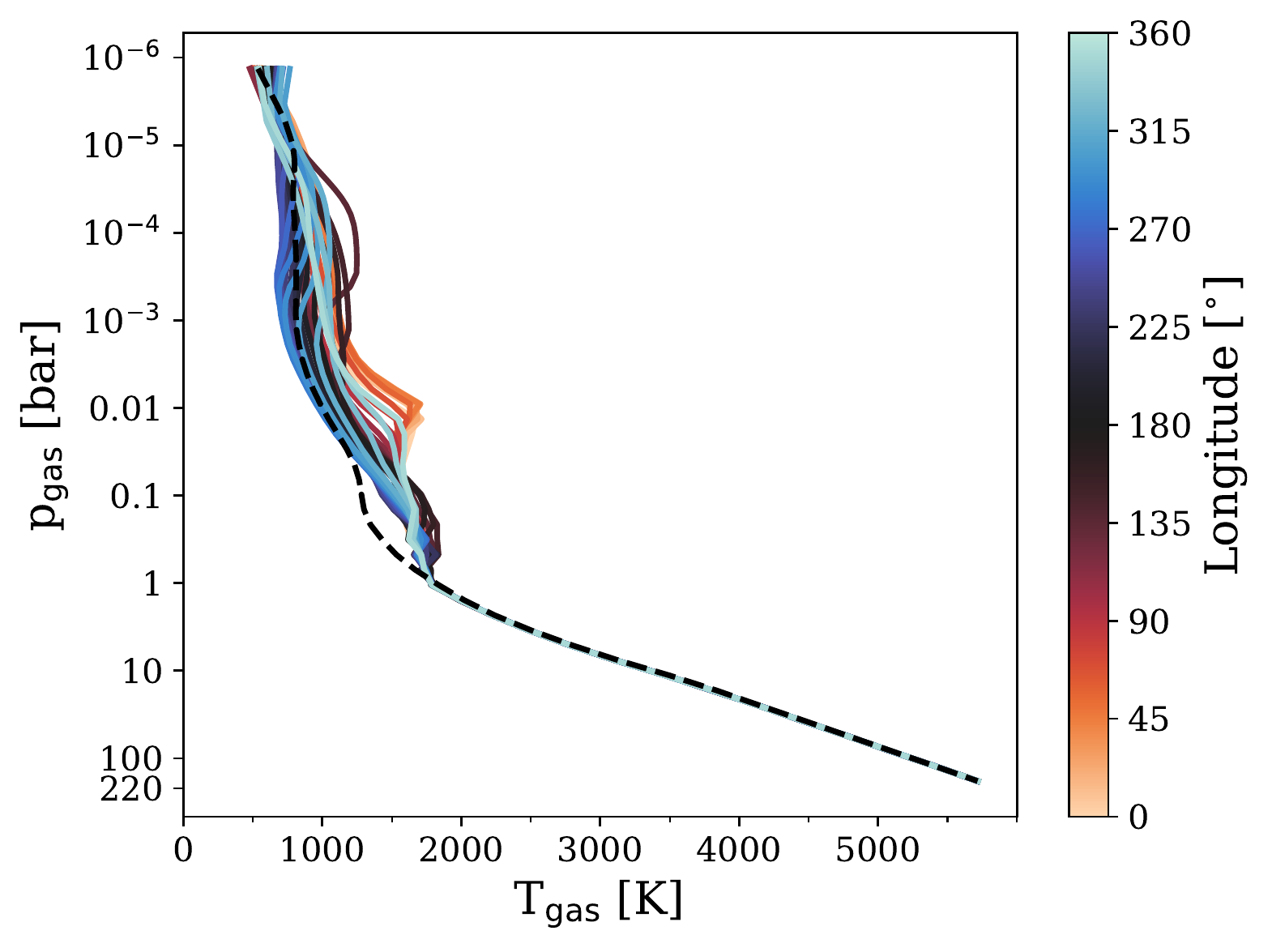}
   \includegraphics[width=0.49\textwidth]{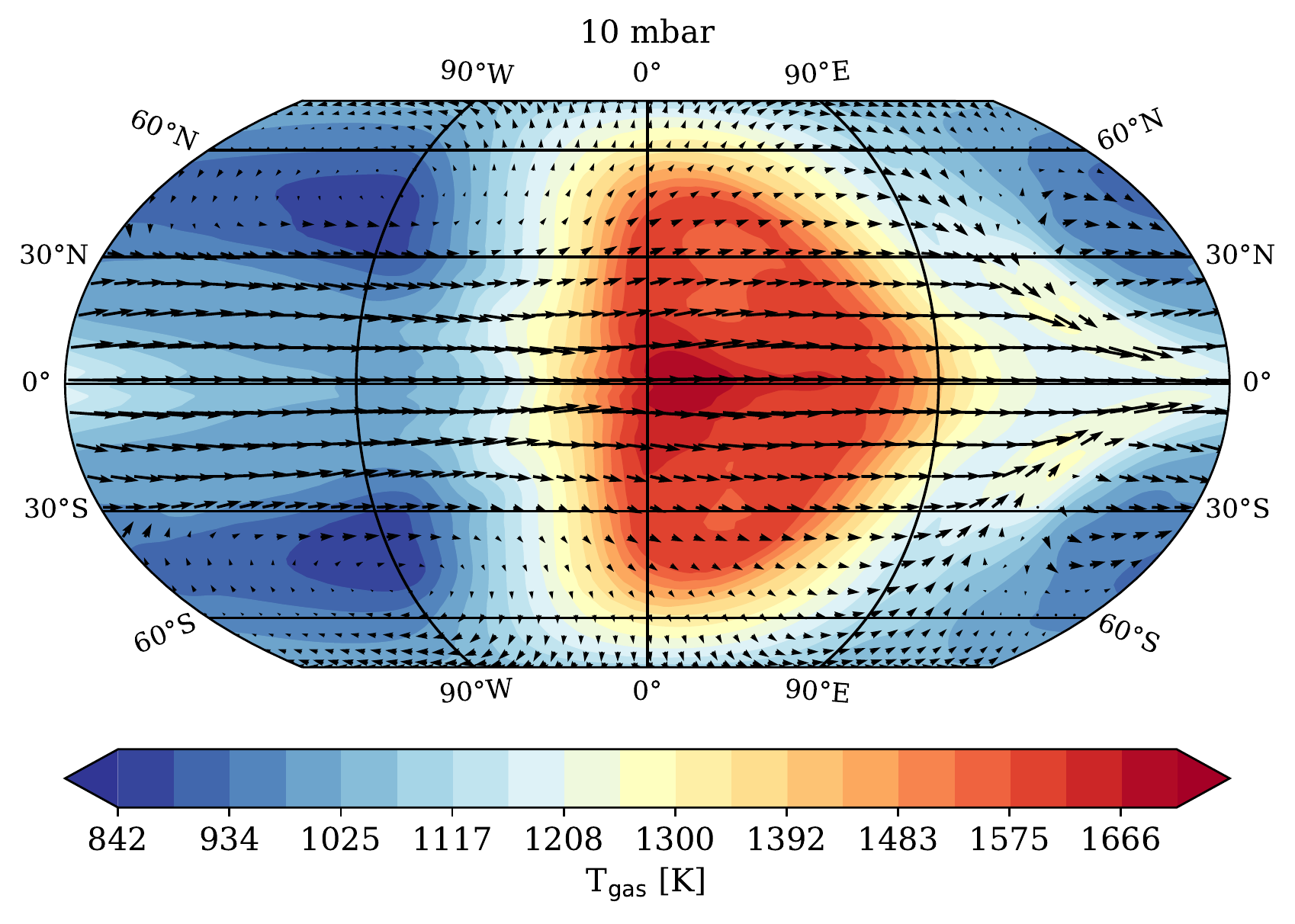}
   \caption{WASP-39b temperature profiles. 
   Left: 1D T-p profiles at the equatorial longitudes (coloured) and polar region (dashed).
   Right: latitude-longitude map at the 10 mbar pressure level of the atmospheric temperature.
   The temperature structures show a typical hot Jupiter pattern, with a temperature hotspot shifted eastward of the sub-stellar point.}
   \label{fig:W39b_temp}
\end{figure*}

\begin{figure*} 
   \centering
   \includegraphics[width=0.32\textwidth]{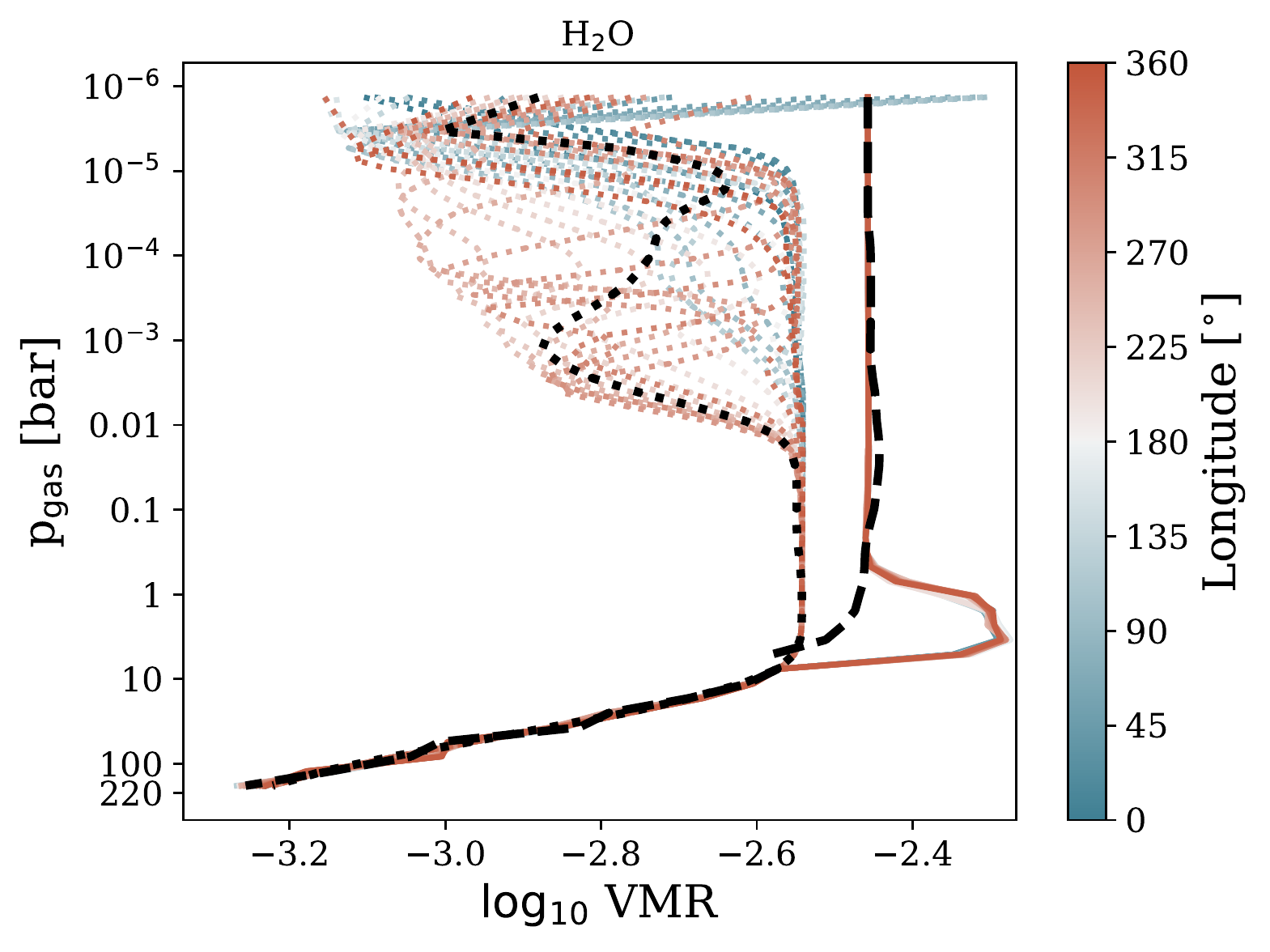}
   \includegraphics[width=0.32\textwidth]{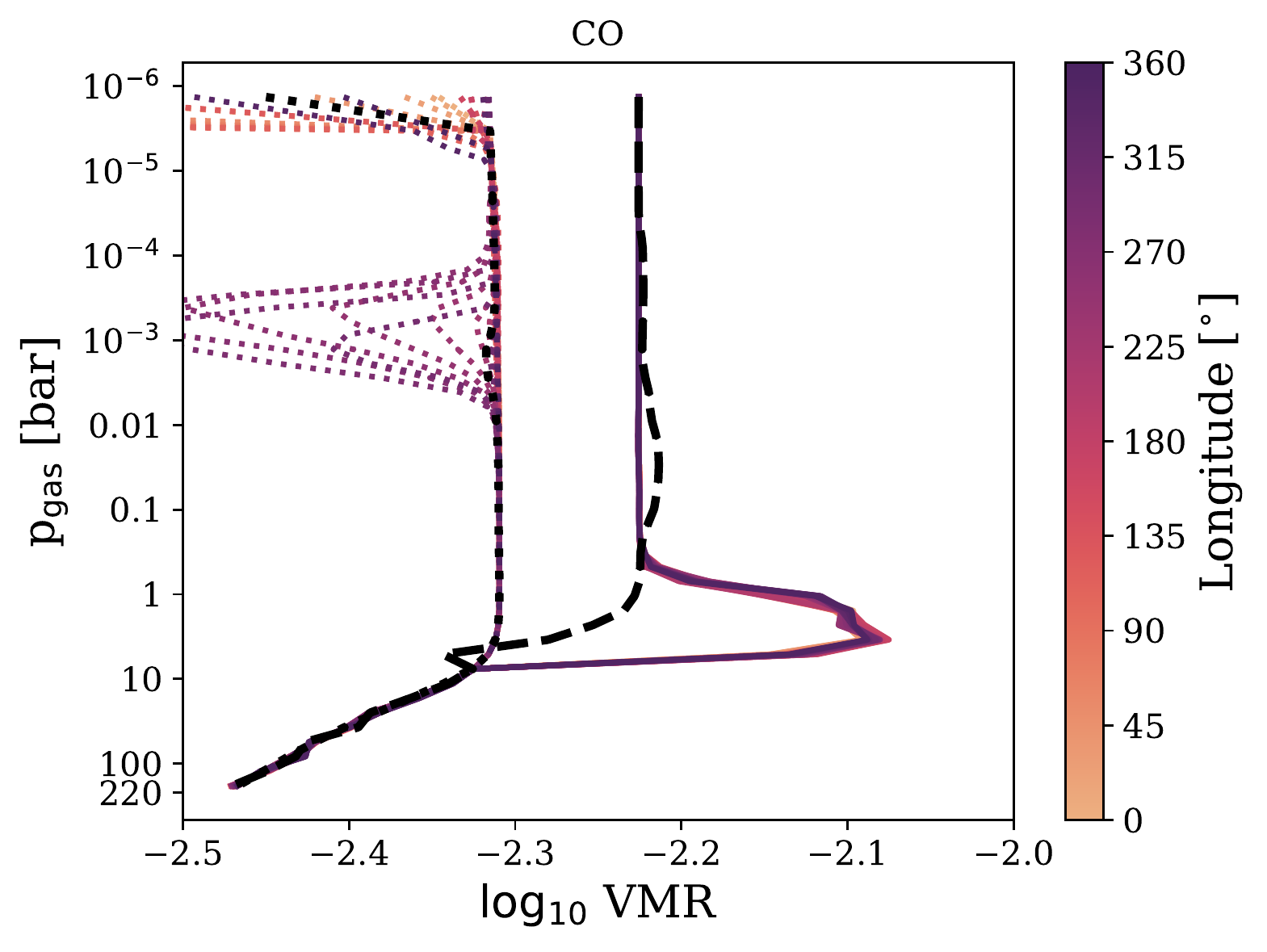}
   \includegraphics[width=0.32\textwidth]{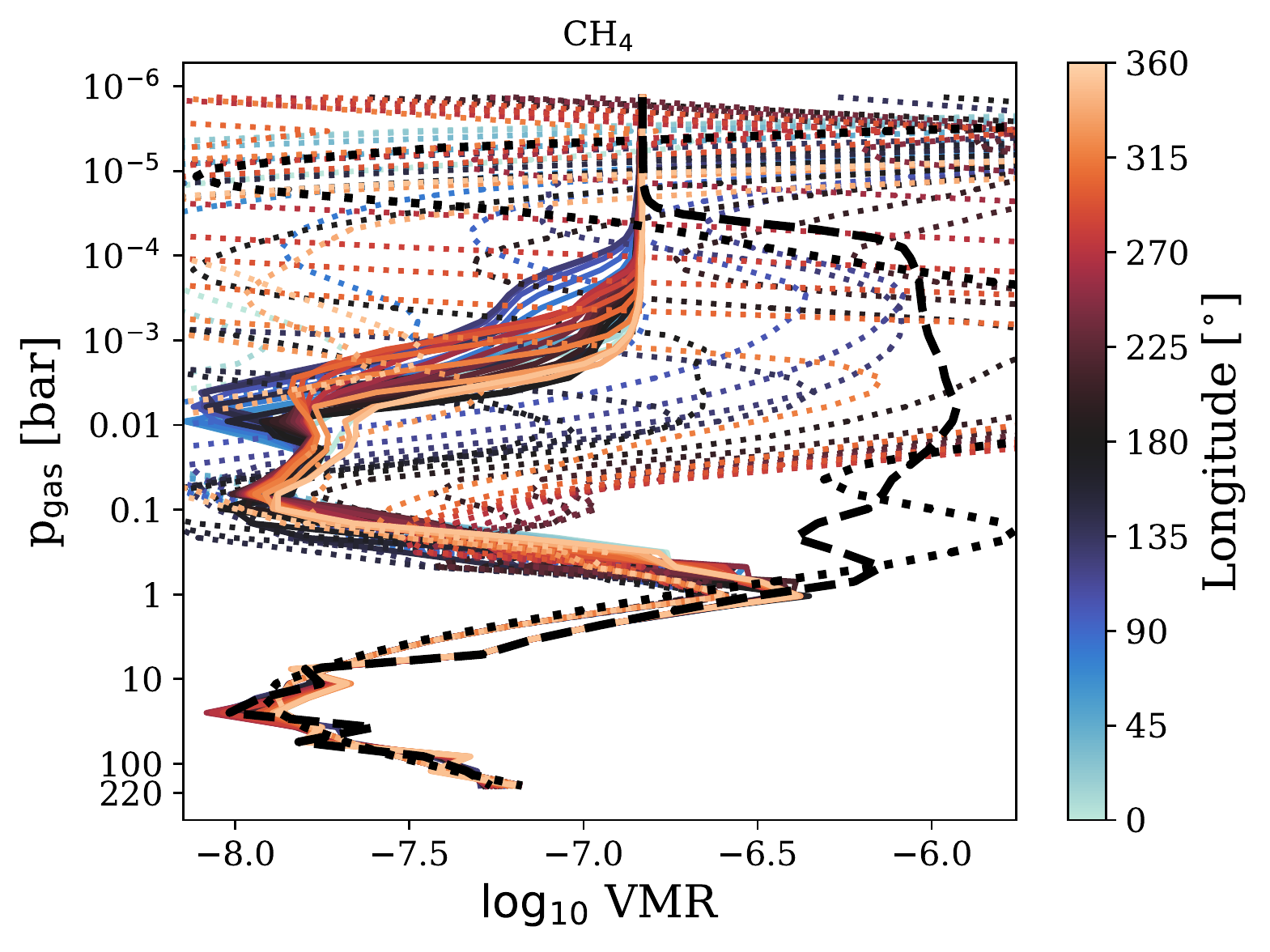}
   \includegraphics[width=0.32\textwidth]{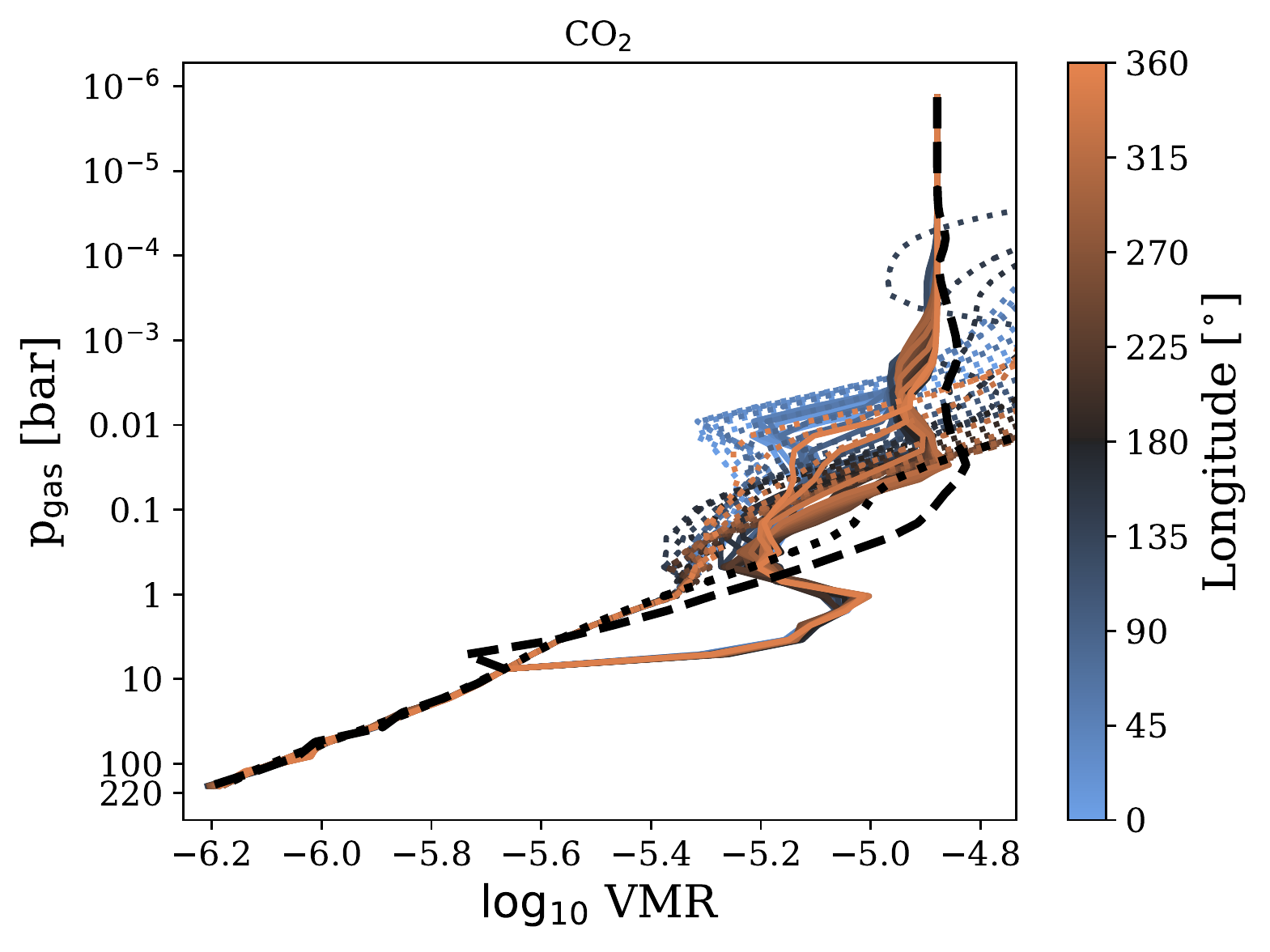}
   \includegraphics[width=0.32\textwidth]{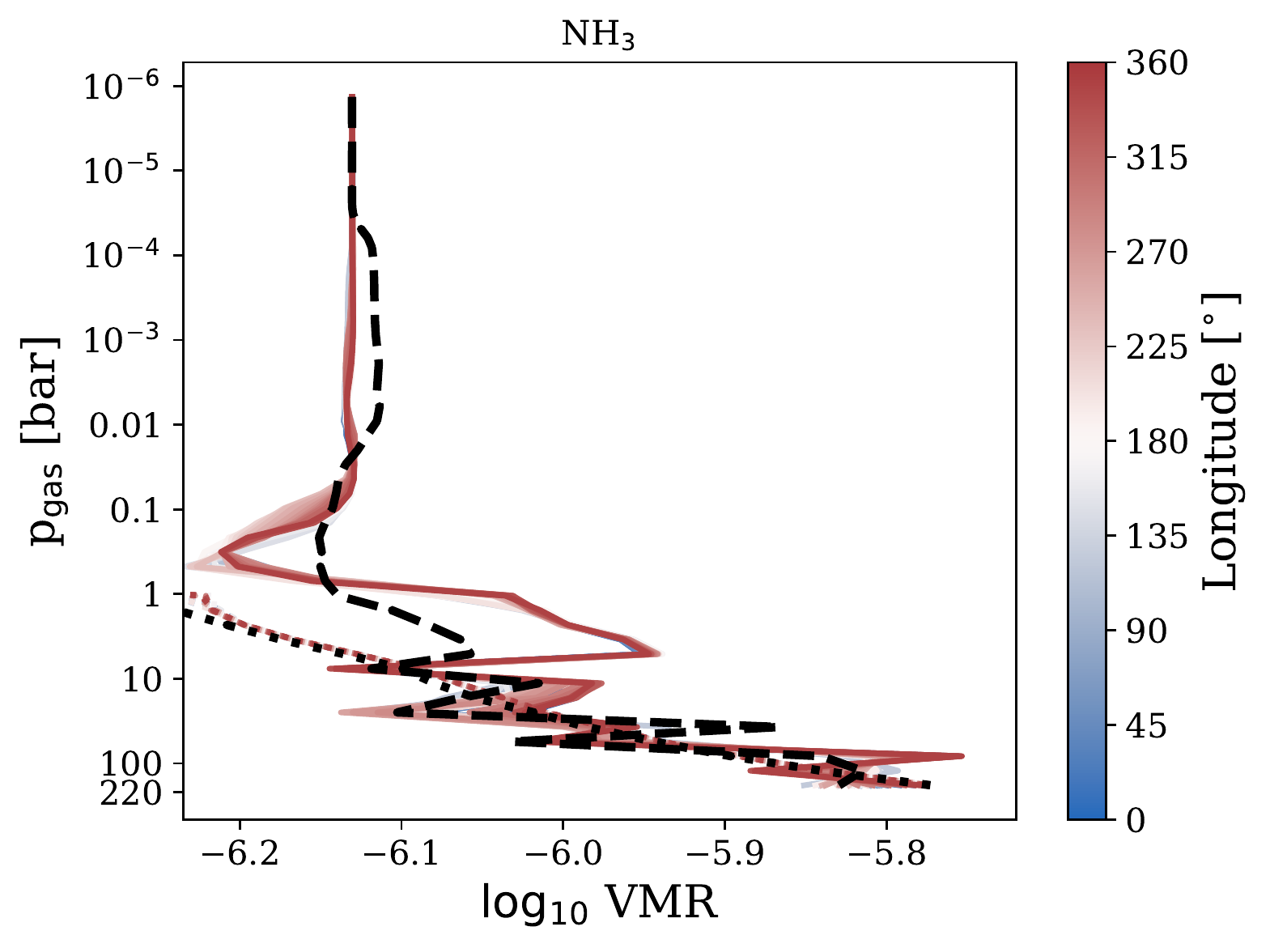}
   \includegraphics[width=0.32\textwidth]{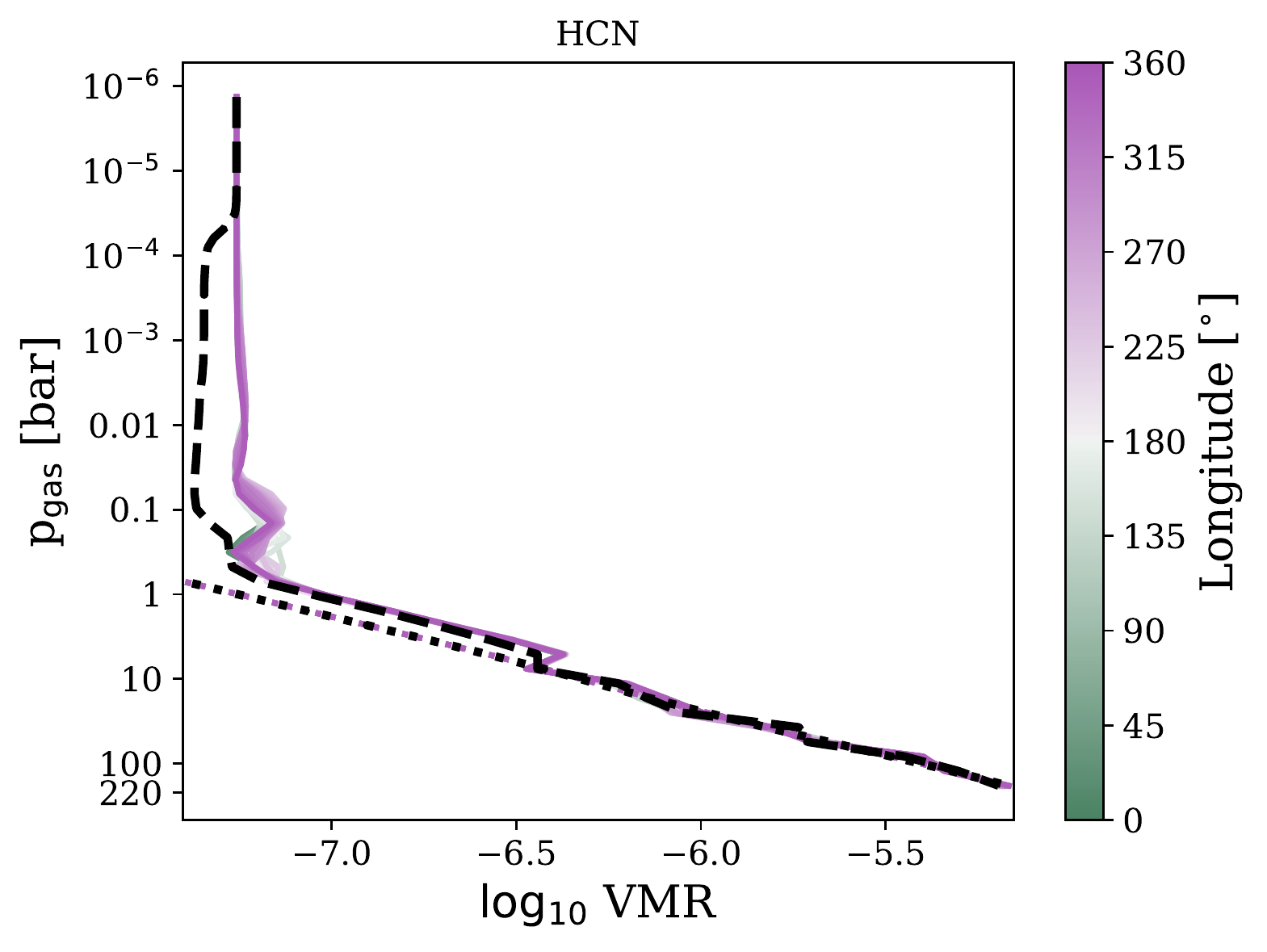}
   \caption{WASP-39b VMR 1D vertical plots. The coloured lines show the variation of VMR with longitude at the equator, while the black dashed line shows the VMR at a polar region. Dotted lines denote the VMR at chemical equilibrium, with the black dotted line denoting the polar region.}
   \label{fig:W39b_vert}
\end{figure*}

\begin{figure*} 
   \centering
   \includegraphics[width=0.45\textwidth]{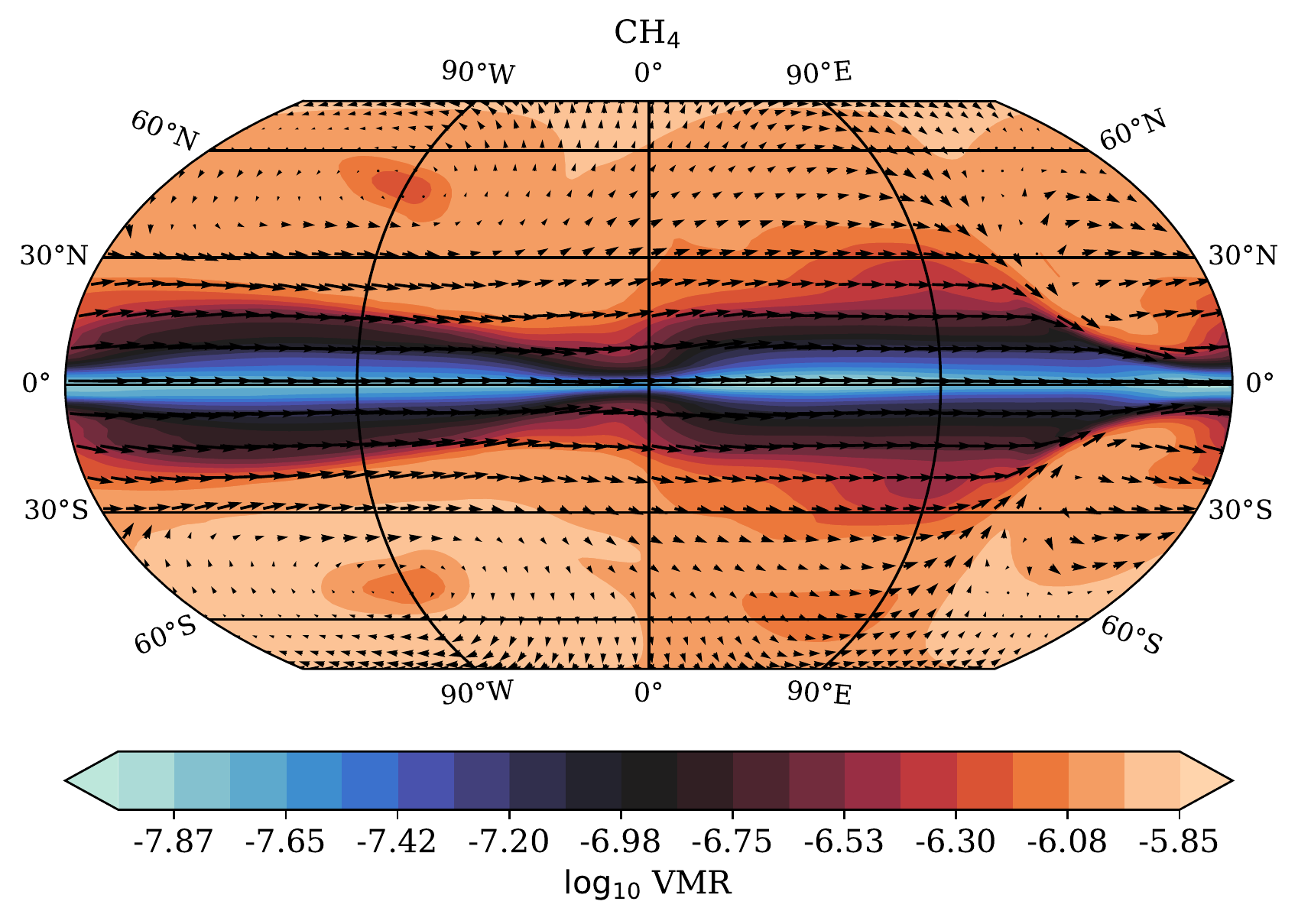}
   \includegraphics[width=0.45\textwidth]{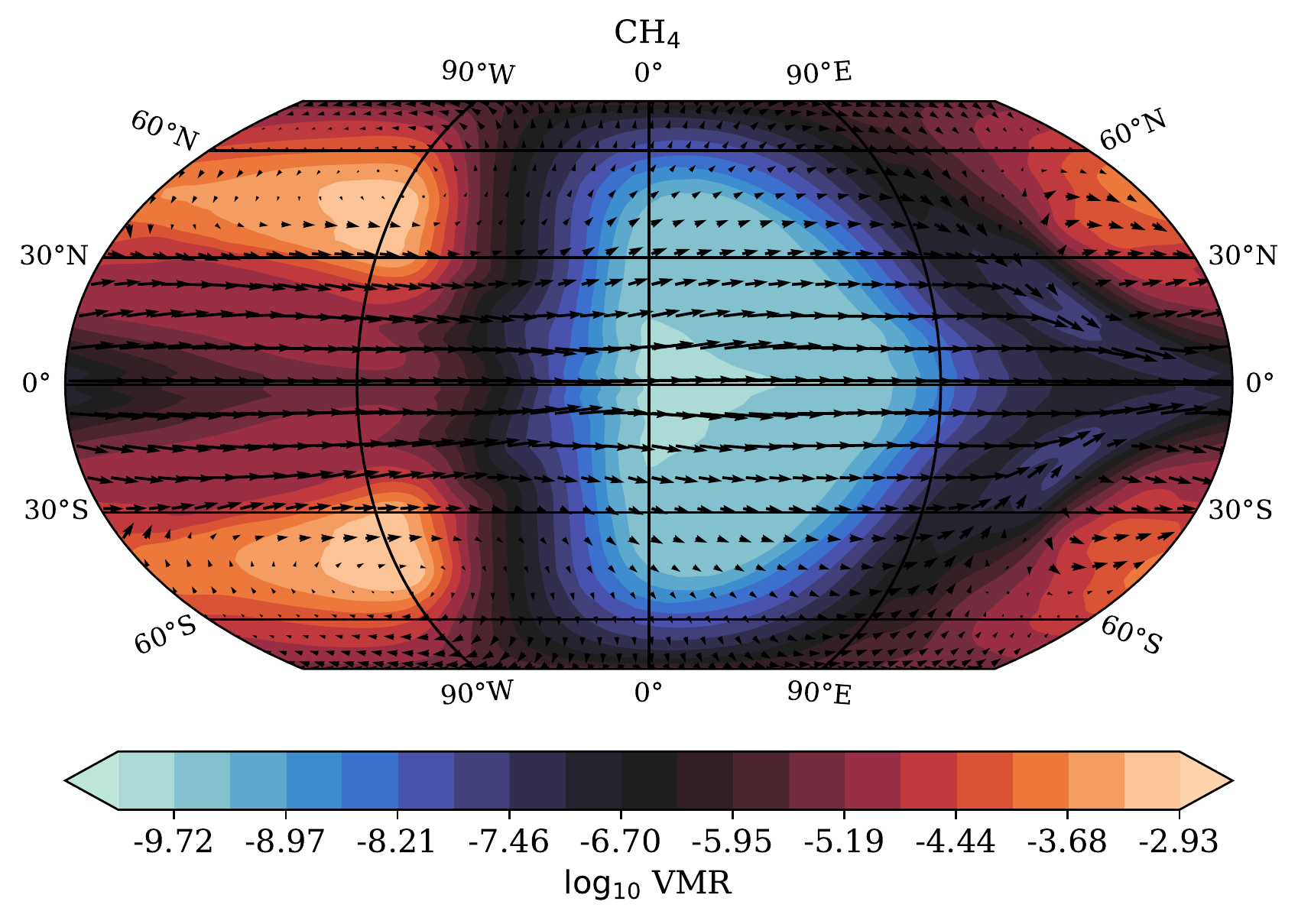}
   \includegraphics[width=0.45\textwidth]{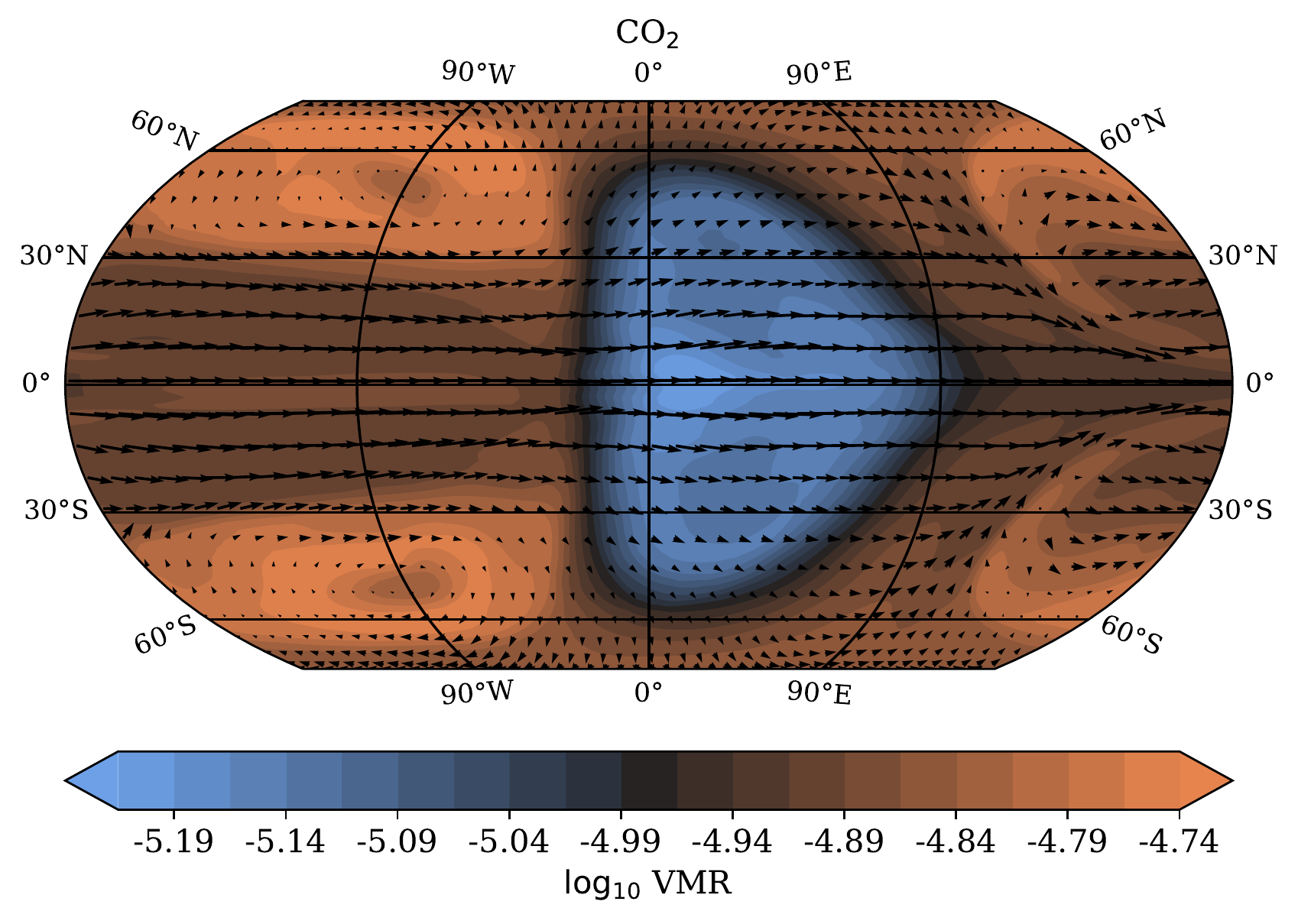}
    \includegraphics[width=0.45\textwidth]{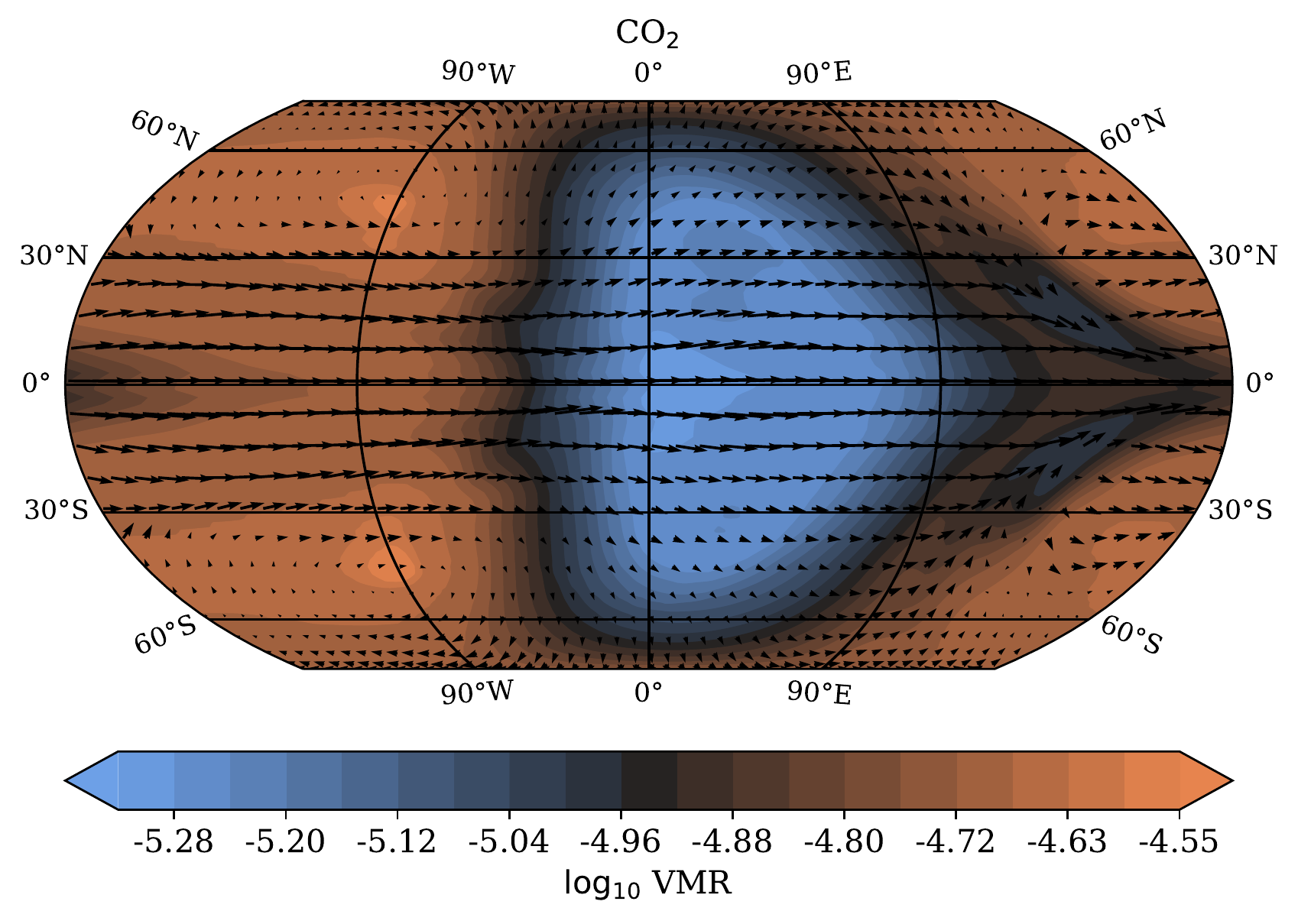}  \includegraphics[width=0.45\textwidth]{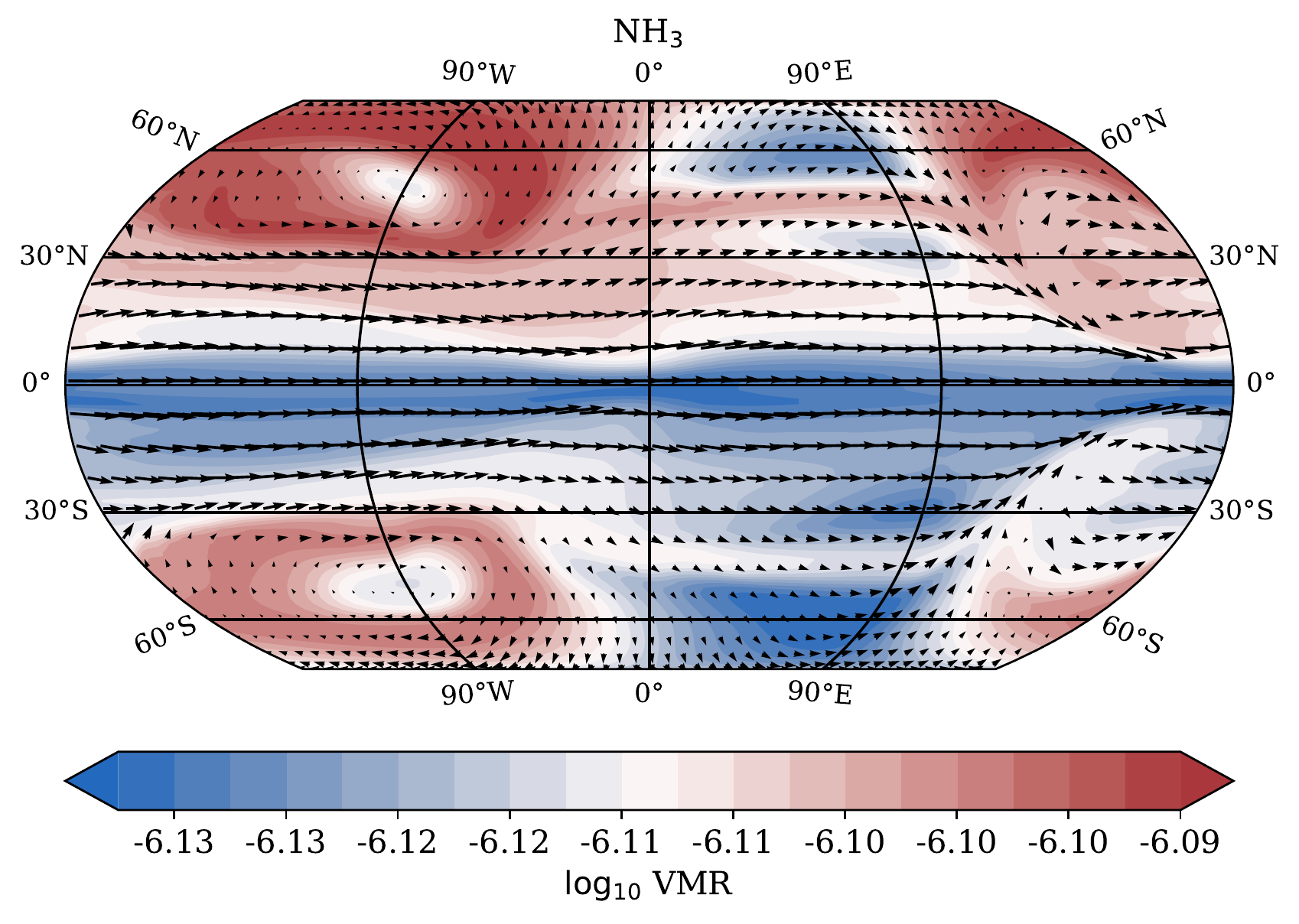}
    \includegraphics[width=0.45\textwidth]{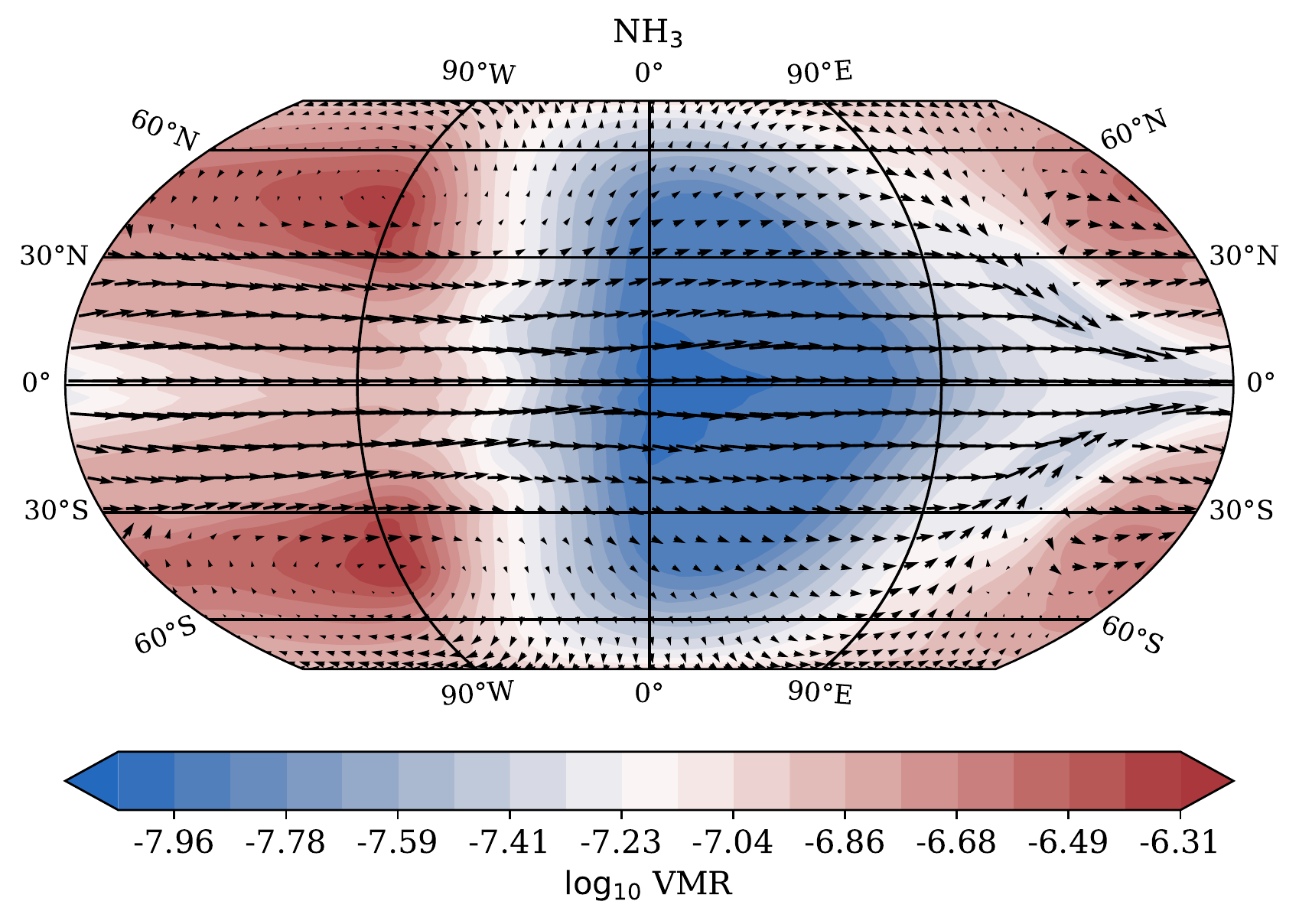}  
    \includegraphics[width=0.45\textwidth]{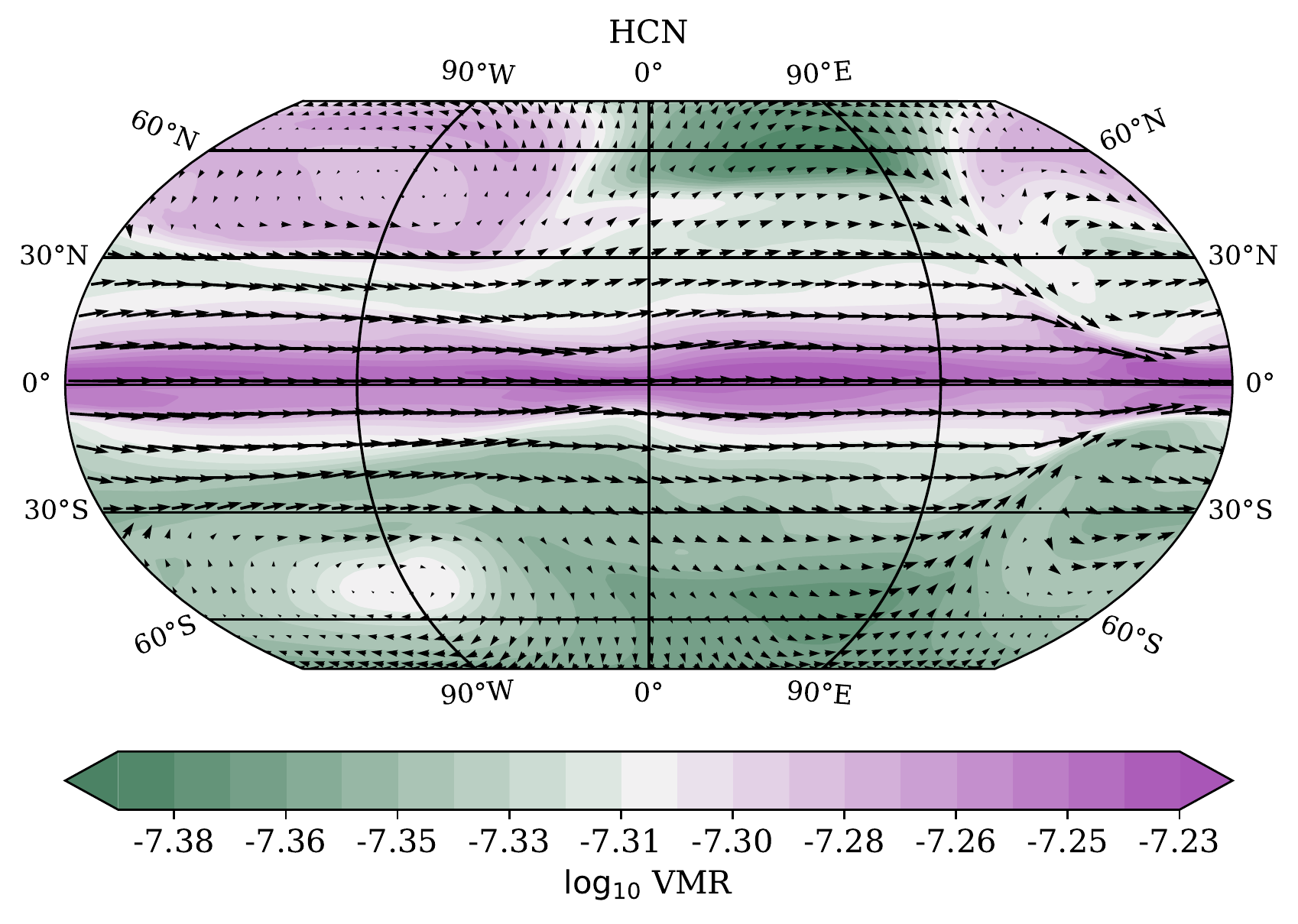}
     \includegraphics[width=0.45\textwidth]{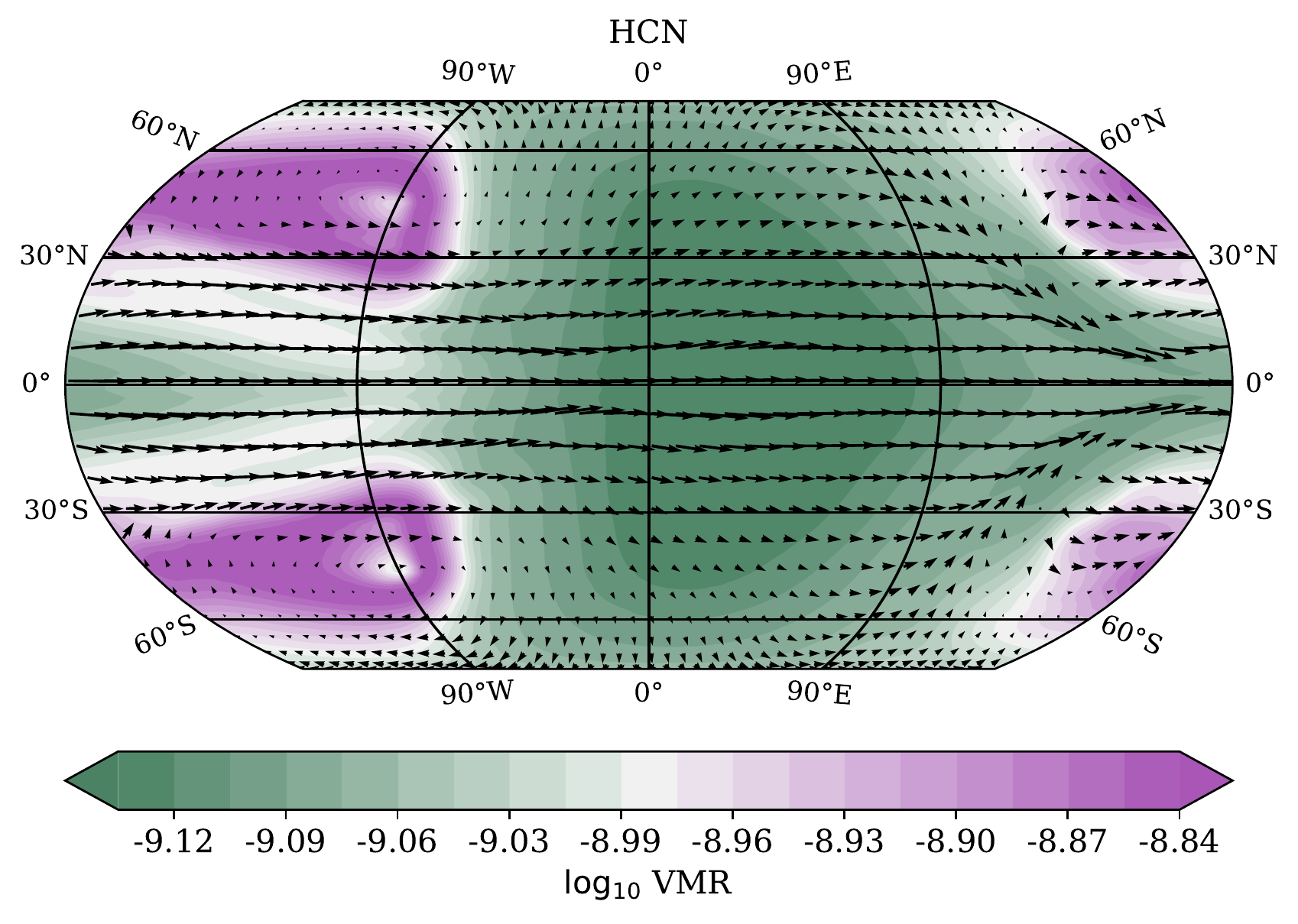}
   \caption{WASP-39b VMR latitude-longitude maps at the 10 mbar pressure level of the atmosphere, showing the variation of VMR with latitude and longitude of the planet.
   Left: results from the mini-chem coupling.
   Right: results assuming chemical equilibrium.}
   \label{fig:W39b_map}
\end{figure*}

In Figure \ref{fig:W39b_temp} we show the WASP-39b vertical temperature-pressure profiles and temperature map at 10 mbar of the GCM simulation.
This shows a typical HJ circulation pattern for this regime \citep[e.g.][]{Showman2020}, with a strong central jet responsible for heat transport around the equatorial regions.
At higher latitude on the western terminator cold Rossby lobes form from the interaction of Rossby waves with the mean flow.

In Figure \ref{fig:W39b_vert} we present the 1D VMR profiles with pressure for CH$_{4}$, CO$_{2}$, NH$_{3}$ and HCN as produced by the mini-chem scheme.
From these plots it is clear that different species have different vertical quench pressures due to their different chemical timescales.
For example, H$_{2}$O, CO, CO$_{2}$ and NH$_{3}$ exhibit quench pressures at around 10 bar, where the kinetic and CE abundances begin to differ significantly, while CH$_{4}$ and HCN quench around 1 bar.
These plots also show the variations between the CE and mini-chem results, with the CE results showing strong dependence on pressure and temperature above the quench pressures, while the mini-chem results are highly homogenised above the quench pressure.
In addition to this, differences in the quench pressure at the polar regions compared to the equatorial are present, in particular for CH$_{4}$ and CO$_{2}$, suggesting that the exact quench pressure is latitude dependent and hence meridional flows also influence the vertical distribution of species.
The VMR is also highly homogenised at the equatorial regions, suggesting that strong zonal quenching is occurring in the simulations.

Our deep atmospheric regions show some non-smooth VMR profiles, especially for CH$_{4}$ and NH$_{3}$. 
This is possibly due to deep dynamical interactions with the lower boundary affecting the vertical advection of tracers through a reflecting wave pattern, or quick timescale temperature adjustments performed by the dry convective adjustment scheme.
A possible solution to this is a vertical homogenisation of tracers across the adiabatic region when the dry convective adjustment scheme is triggered.
Another possibility is numerical instability through the net reaction CH$_{4}$ + NH$_{3}$ $\rightarrow$ HCN + 3H$_{2}$ for high pressure regions that are near or at chemical equilibrium. This is a known problem from 1D modelling efforts \citep[e.g.][]{Rimmer2016,Tsai2017}, possibly indicating that a stricter criteria to detect (near) chemical equilibrium is required for the model.
However, these variations are small in magnitude overall and only affect the deep regions well below the photosphere of the planet.

In Figure \ref{fig:W39b_map} we show the latitude-longitude VMR map of our chosen four species at the 10 mbar pressure level of the GCM.
We also show the VMR assuming CE at the same pressure levels.
Our results demonstrate that a significant departure from the CE distribution of species, which is highly correlated with the temperature of the atmosphere, between the mini-chem results is seen.
Our mini-chem results are more suggestive that the chemical structure is more thermally and dynamically driven, with species differentiated in VMR between the high latitude wave driven dynamical regime and the equatorial jet latitudes.
Interesting decreases of the VMR at the equatorial region for CH$_{4}$ are seen, suggesting strong homogenisation but also enhancement of the VMR of species across the equatorial section.
HCN is enhanced at the equatorial regions.
Strong homogenisation is also seen in the \citet{Drummond2020} study, but they find a decrease of the HCN VMR at the equatorial areas rather than an increase.
This is possibly due to the different metallicity and thermal structure of the WASP-39b models compared to HD 189733b.
Strong equatorial homogenisation is also seen in pseudo-2D studies that use equatorial T-p and wind profiles derived from GCM models \citep[e.g.][]{Baeyens2021}.

\subsection{HD 189733b}

\begin{figure*} 
   \centering
   \includegraphics[width=0.49\textwidth]{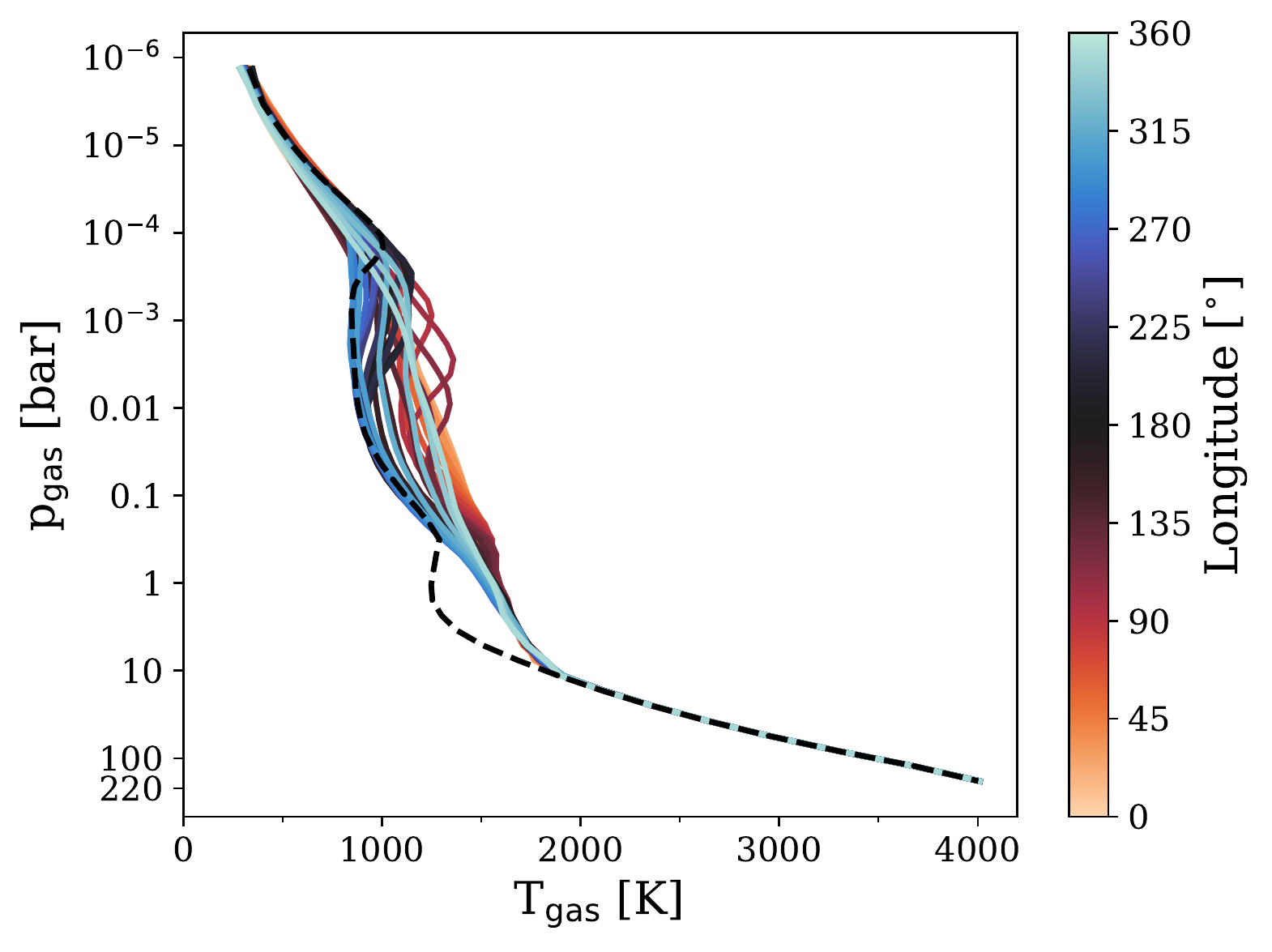}
   \includegraphics[width=0.49\textwidth]{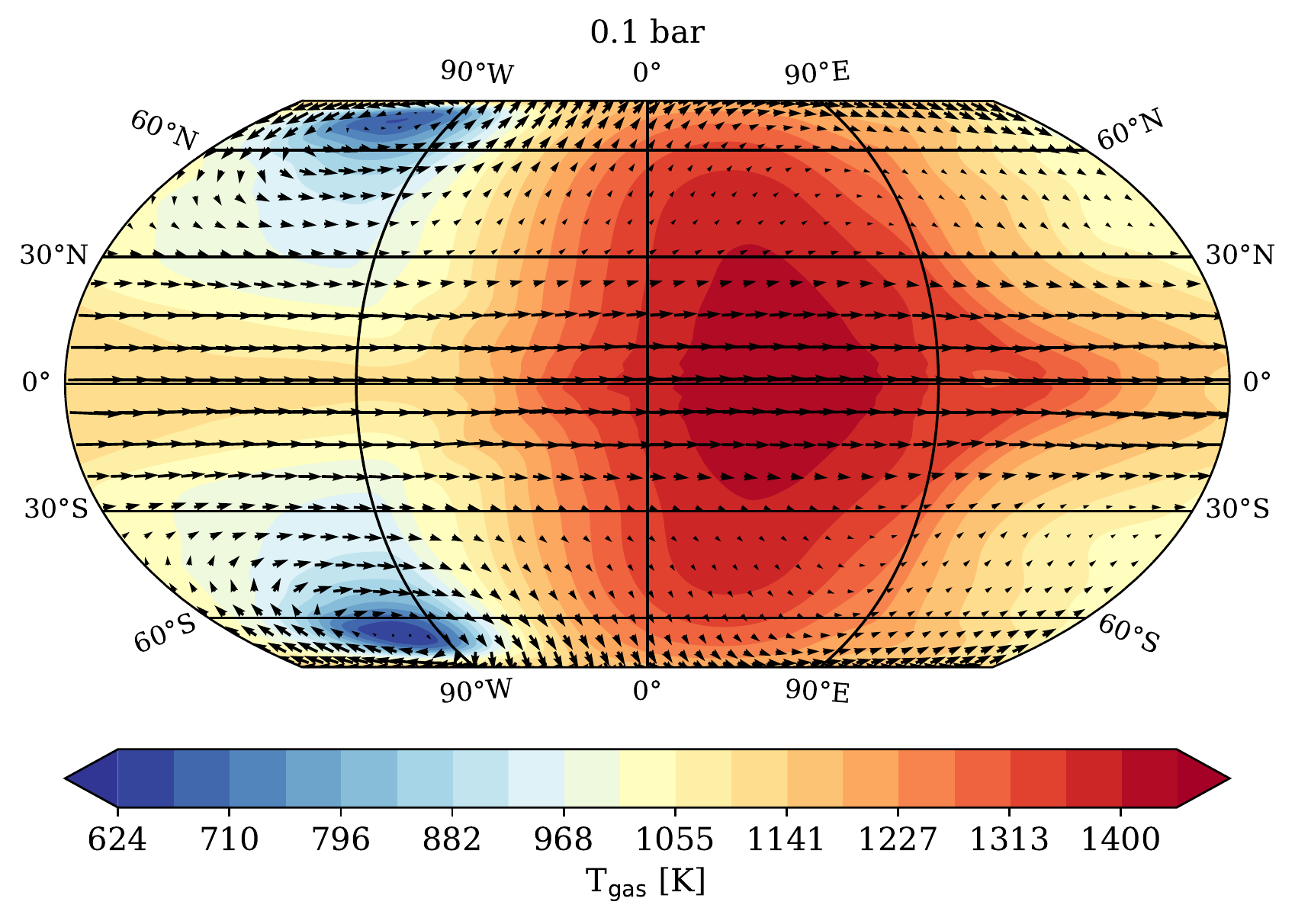}
   \caption{HD 189733b temperature profiles. 
   Left: 1D T-p profiles at the equatorial longitudes (coloured) and polar region (dashed).
   Right: latitude-longitude map at the 0.1 bar pressure level of the atmospheric temperature.}
   \label{fig:HD189b_temp}
\end{figure*}

\begin{figure*} 
   \centering
   \includegraphics[width=0.32\textwidth]{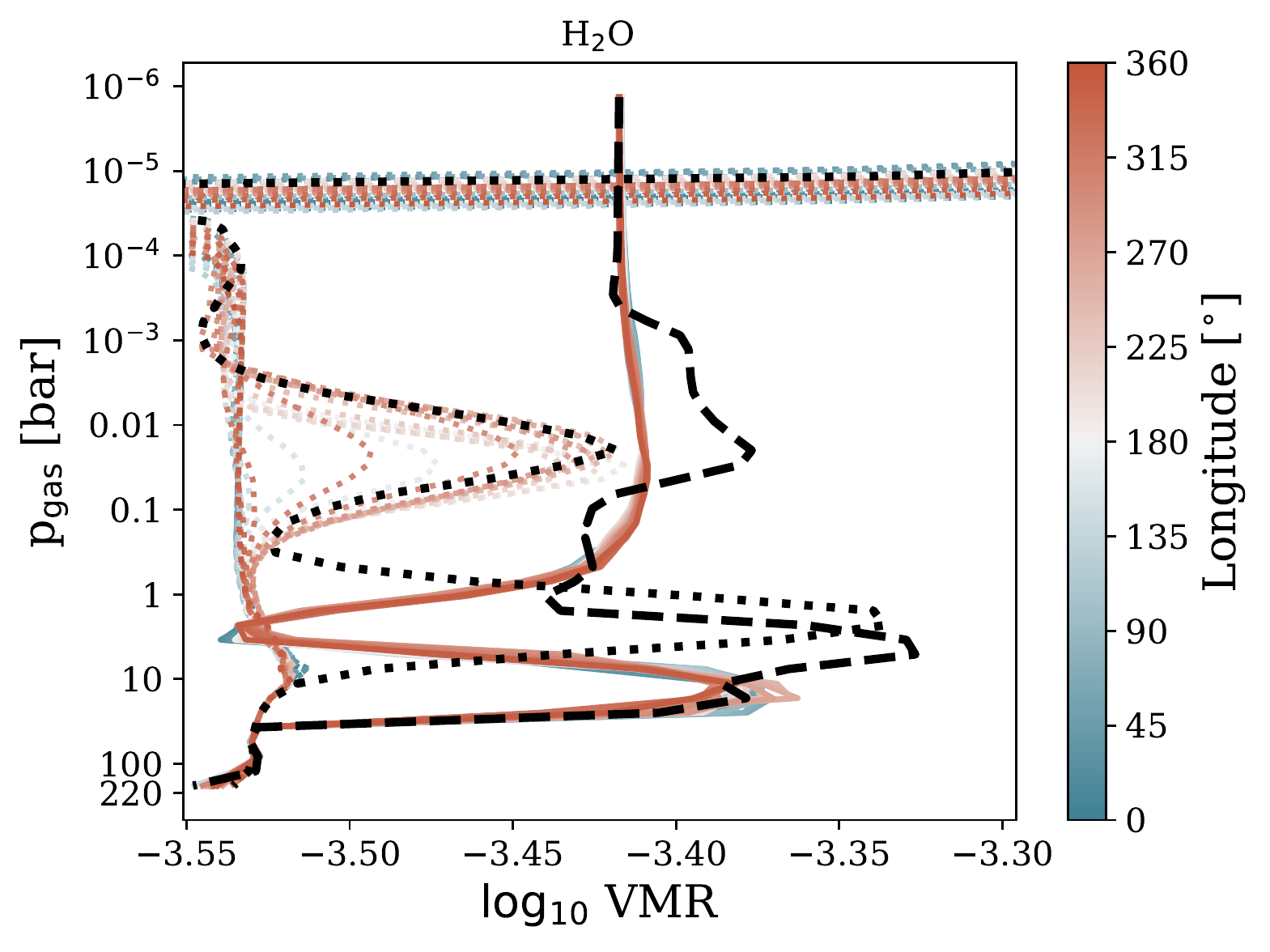}
   \includegraphics[width=0.32\textwidth]{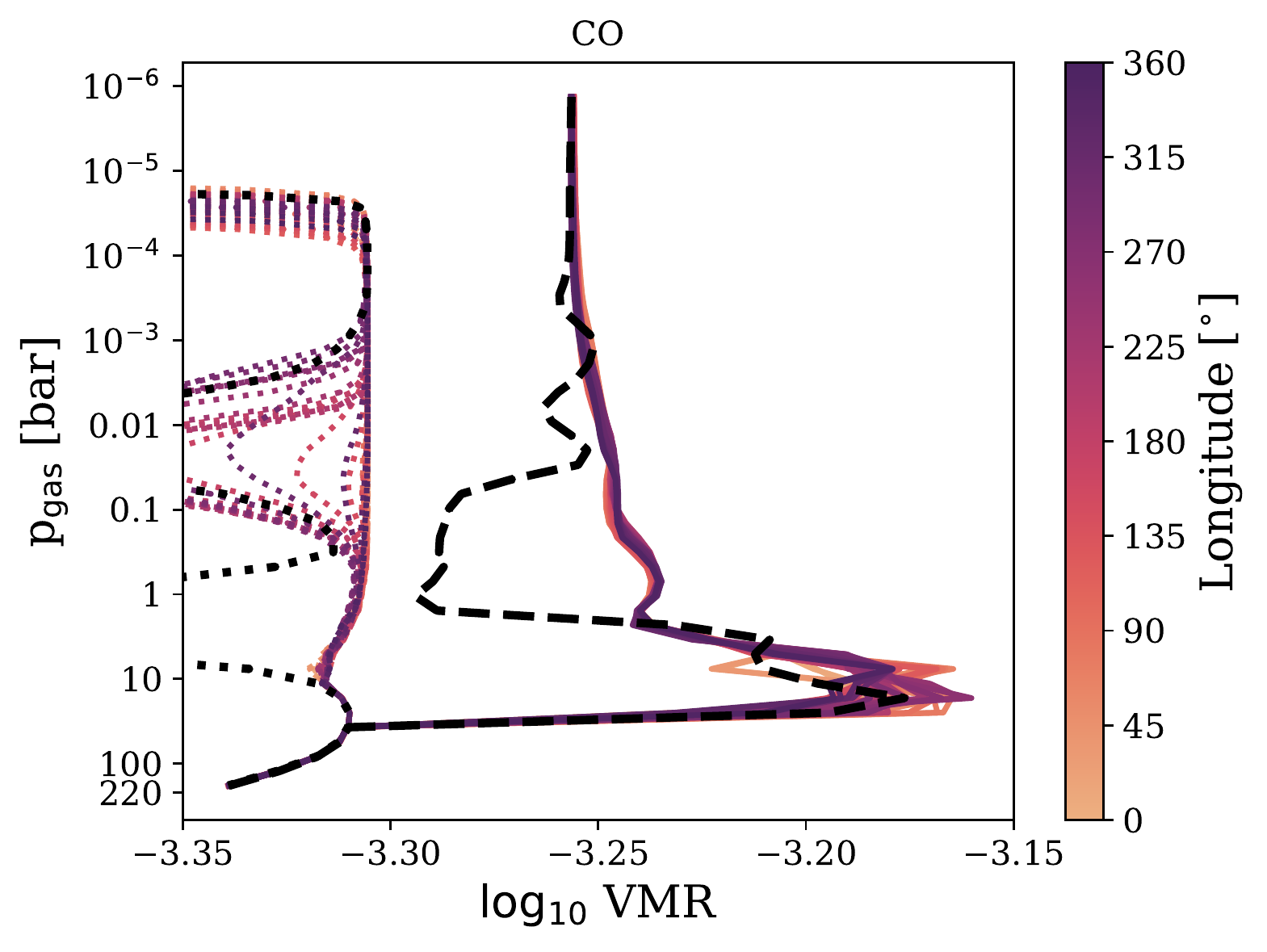}
   \includegraphics[width=0.32\textwidth]{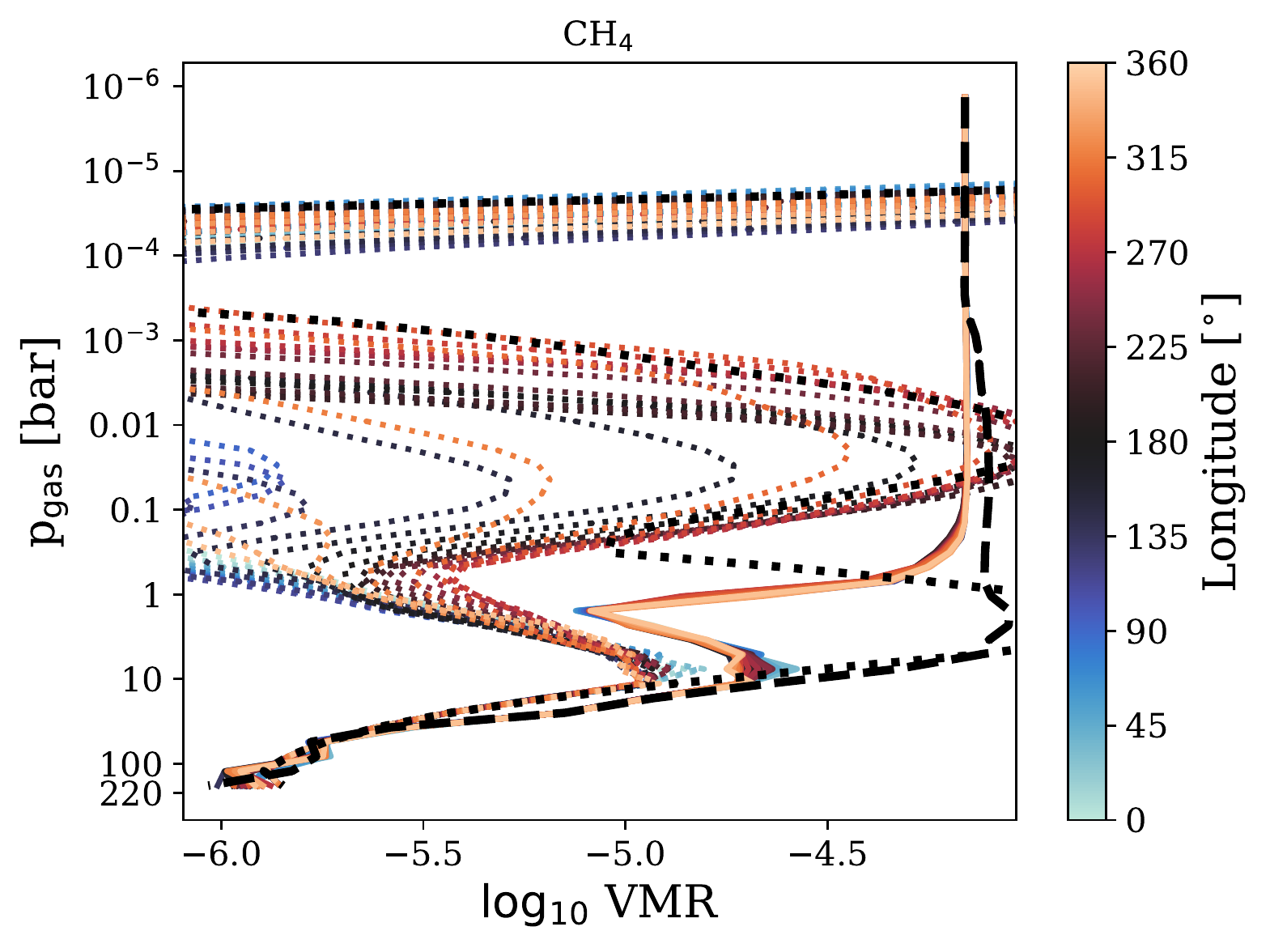}
   \includegraphics[width=0.32\textwidth]{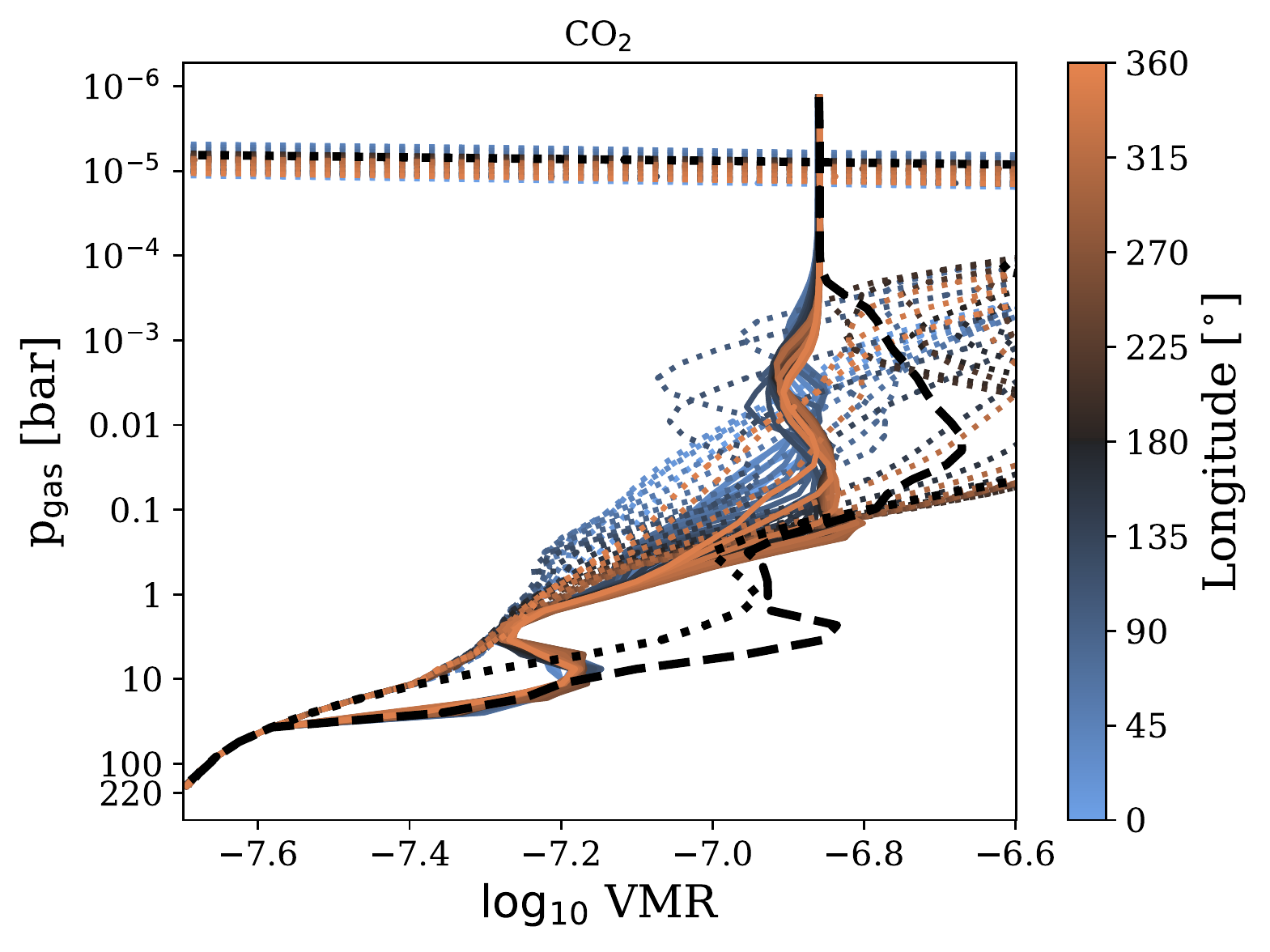}
   \includegraphics[width=0.32\textwidth]{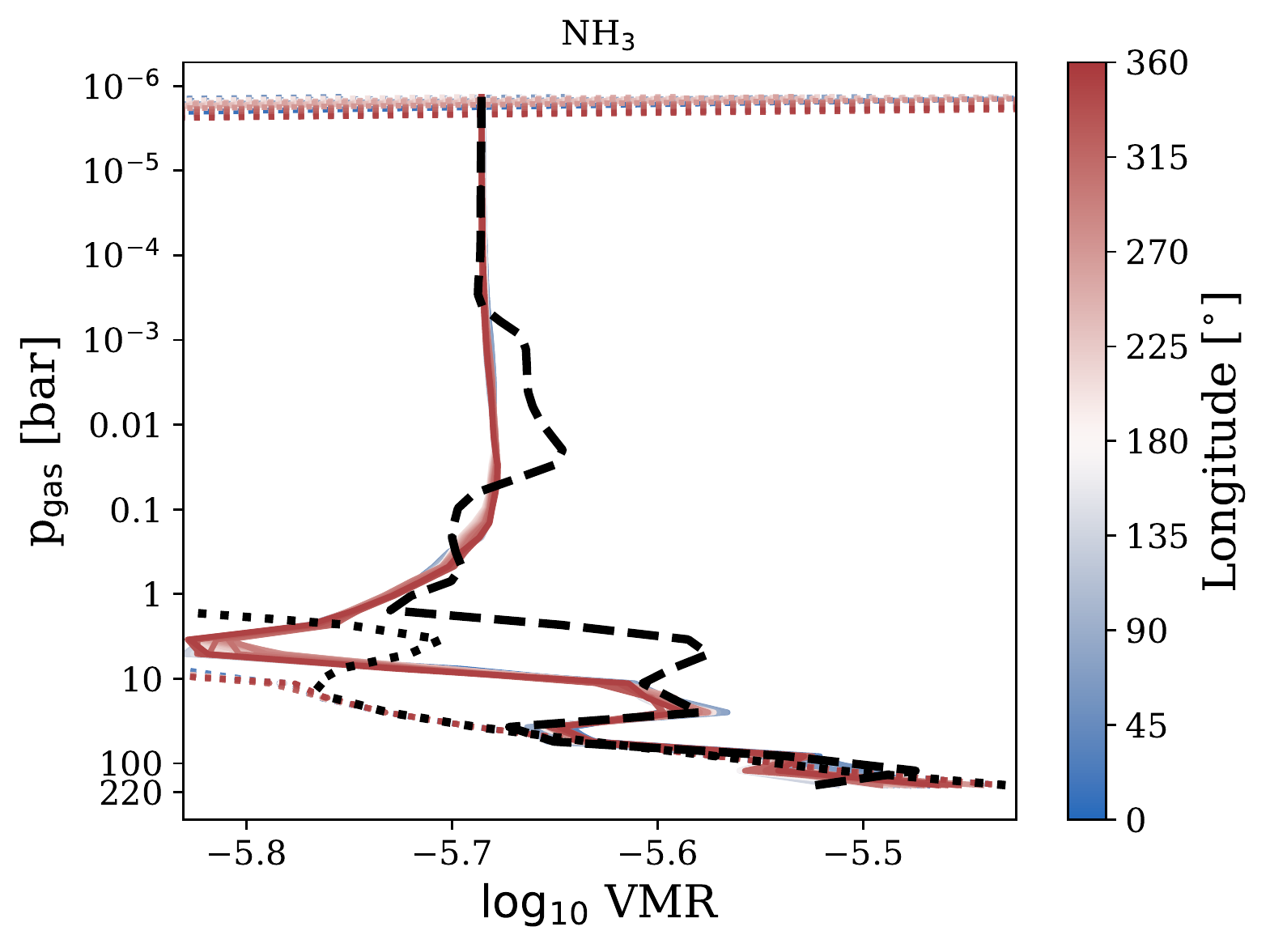}
   \includegraphics[width=0.32\textwidth]{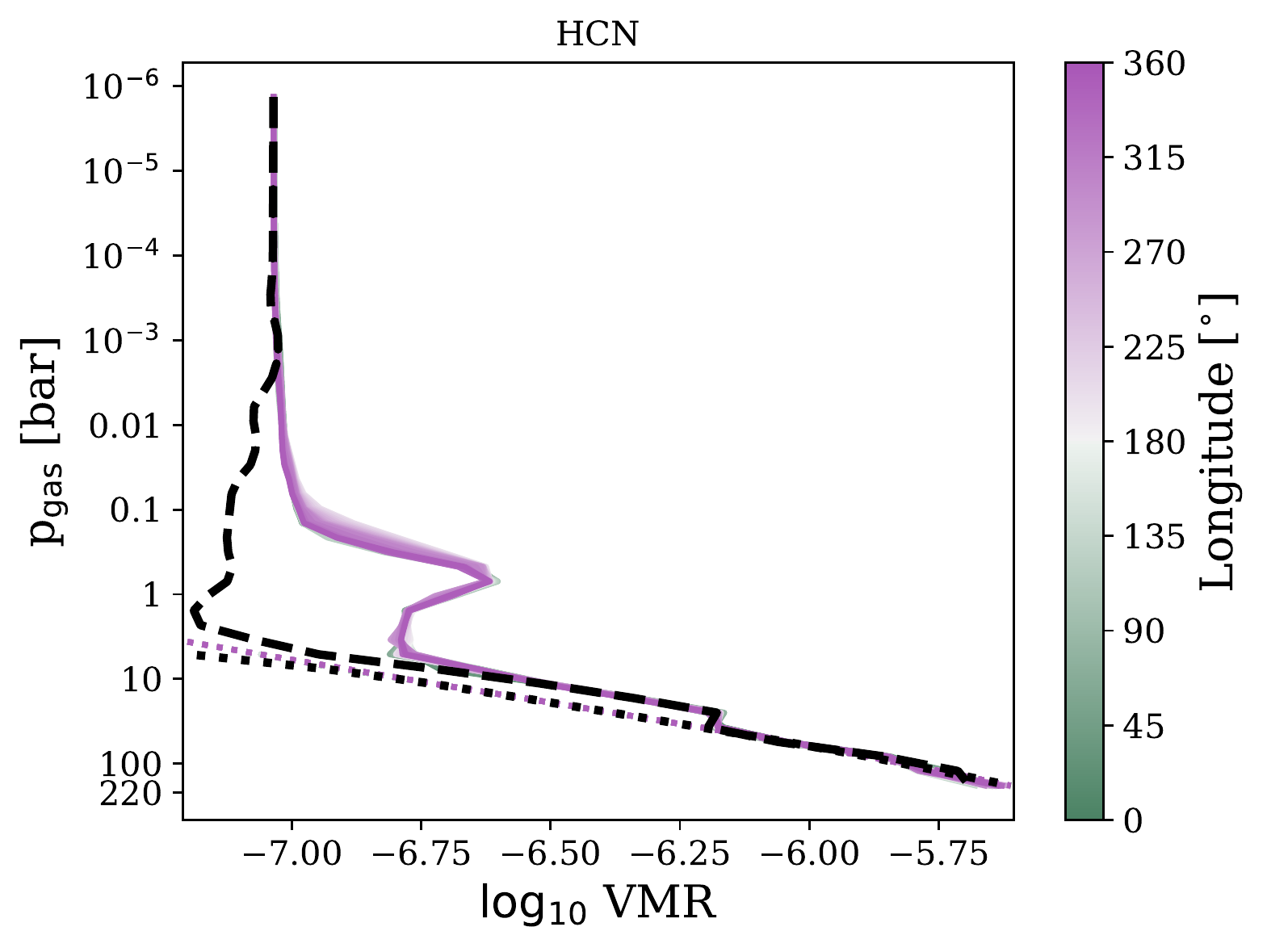}
   \caption{HD 189733b VMR 1D vertical plots. The coloured lines show the variation of VMR with longitude at the equator, while the dashed line shows the VMR at a polar region. Dotted lines denote the VMR at chemical equilibrium, with the black dotted line denoting the polar region.}
   \label{fig:HD189b_vert}
\end{figure*}

\begin{figure*} 
   \centering
   \includegraphics[width=0.45\textwidth]{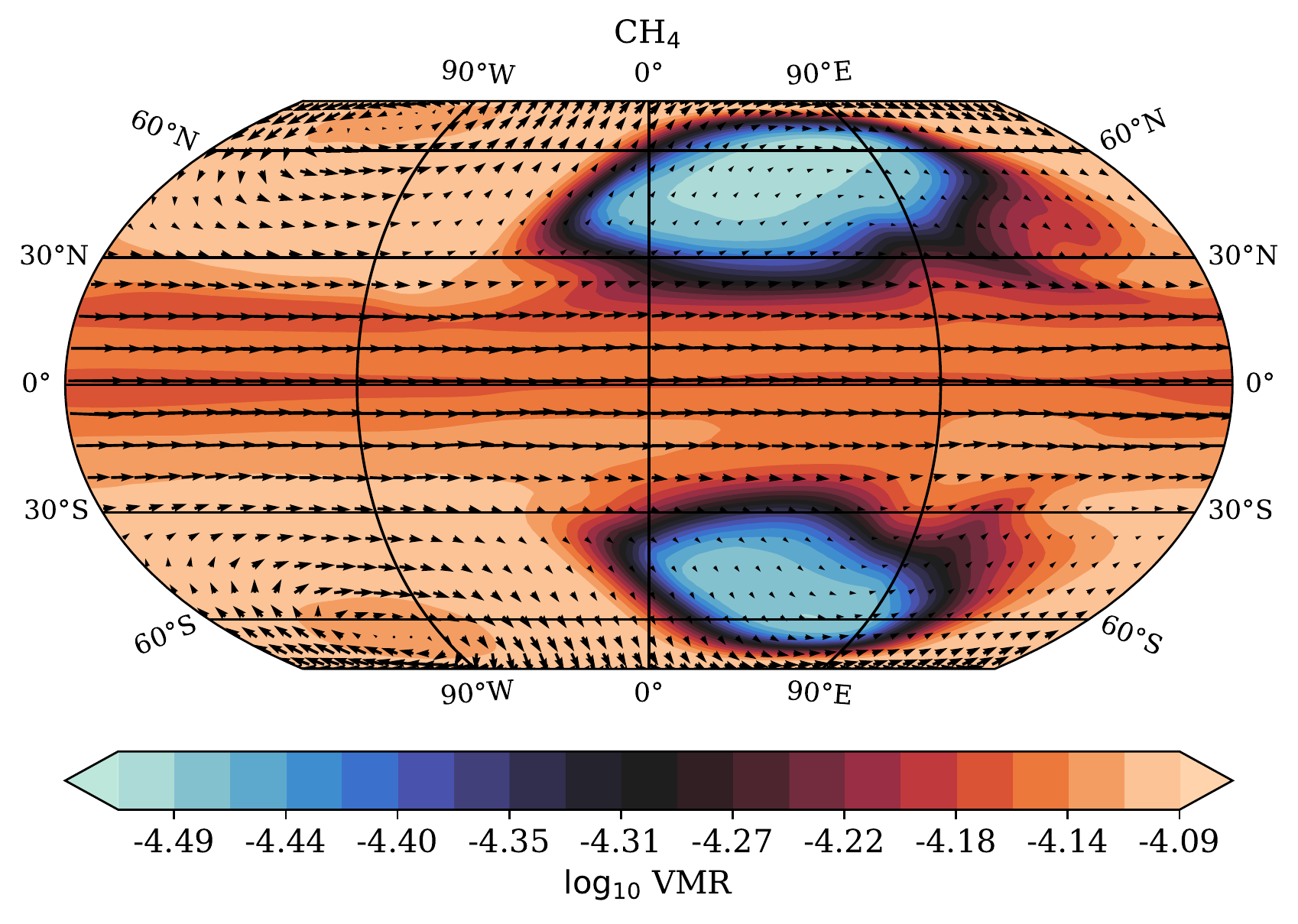}
   \includegraphics[width=0.45\textwidth]{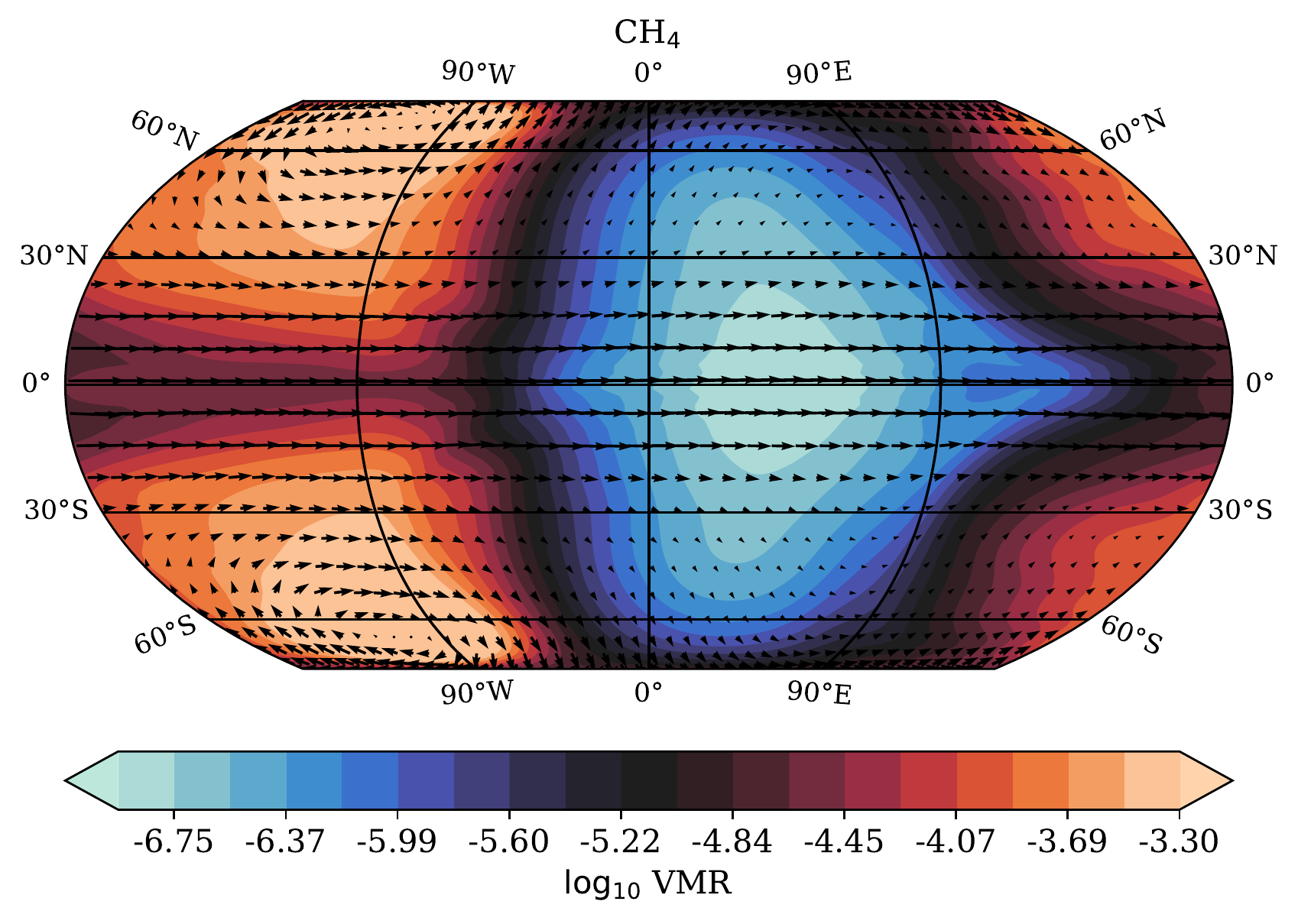}
   \includegraphics[width=0.45\textwidth]{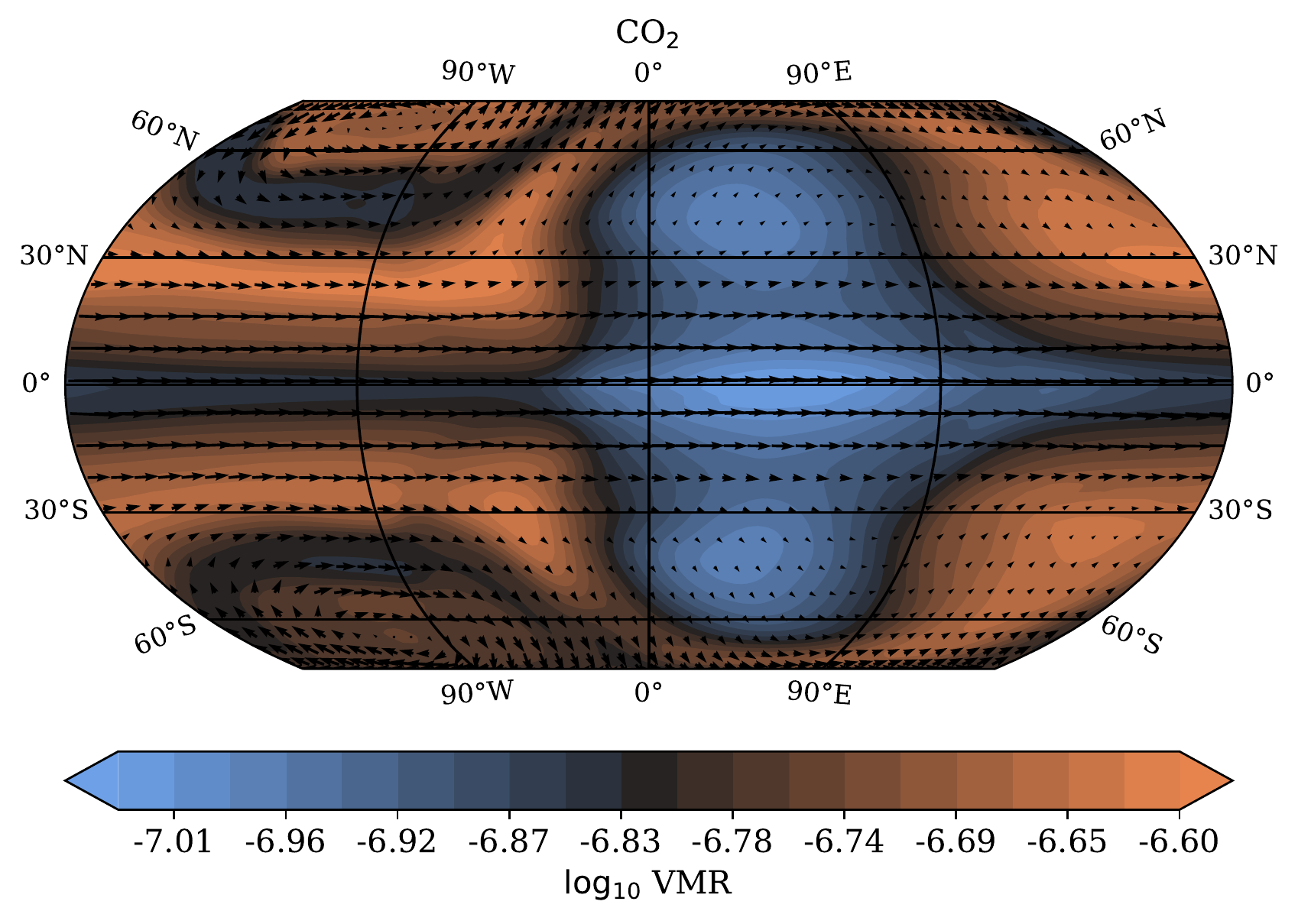}
   \includegraphics[width=0.45\textwidth]{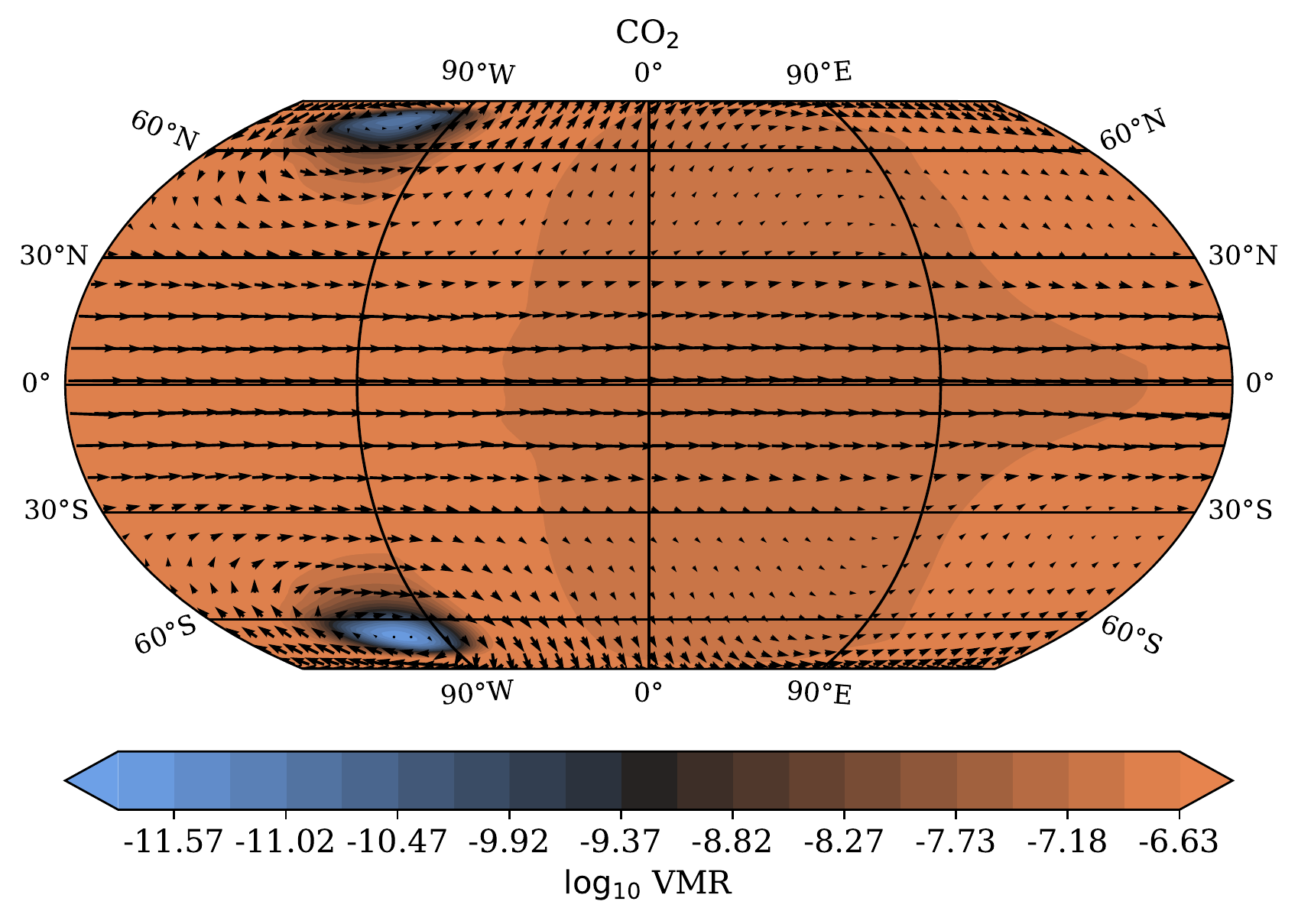}
   \includegraphics[width=0.45\textwidth]{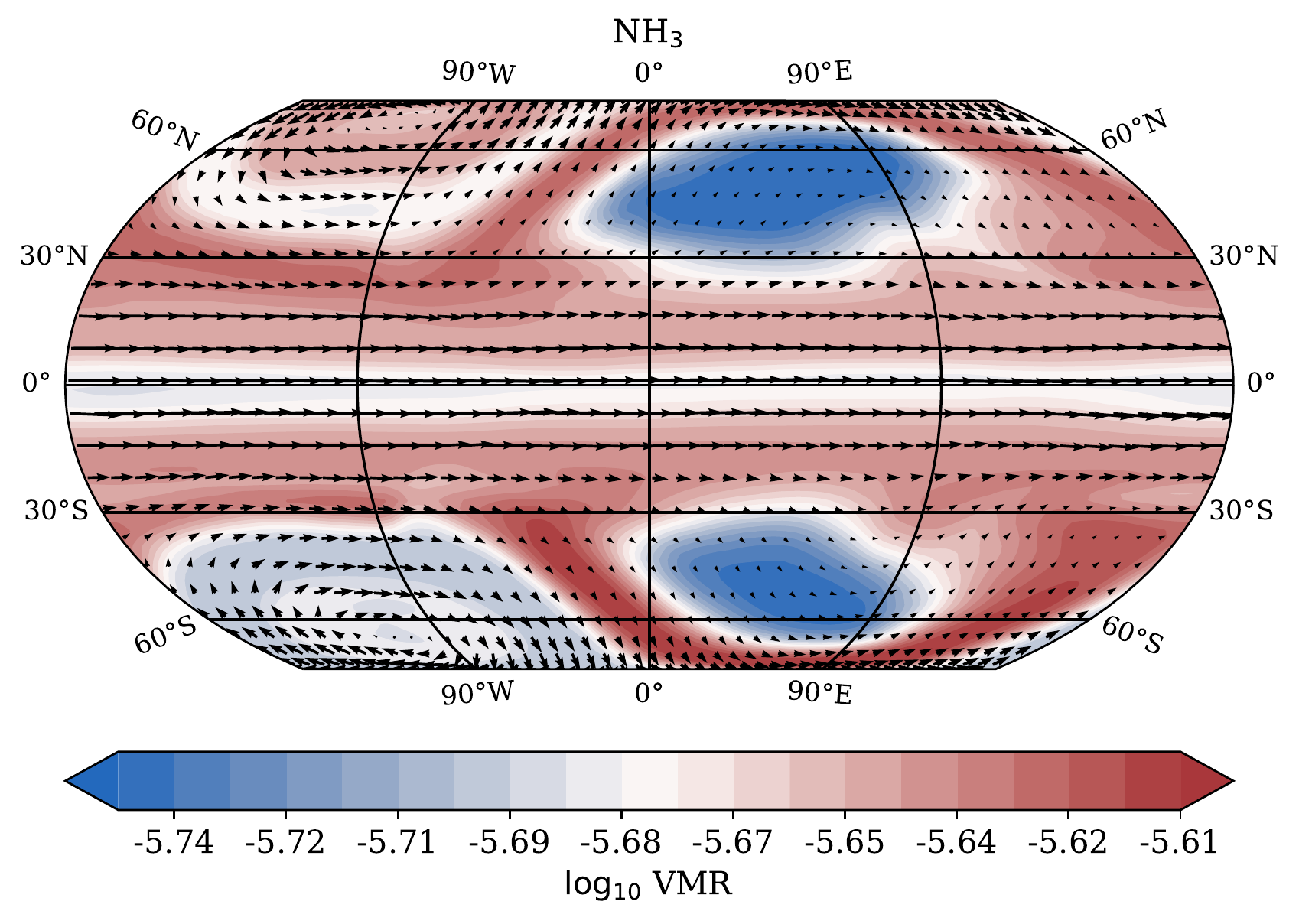}
   \includegraphics[width=0.45\textwidth]{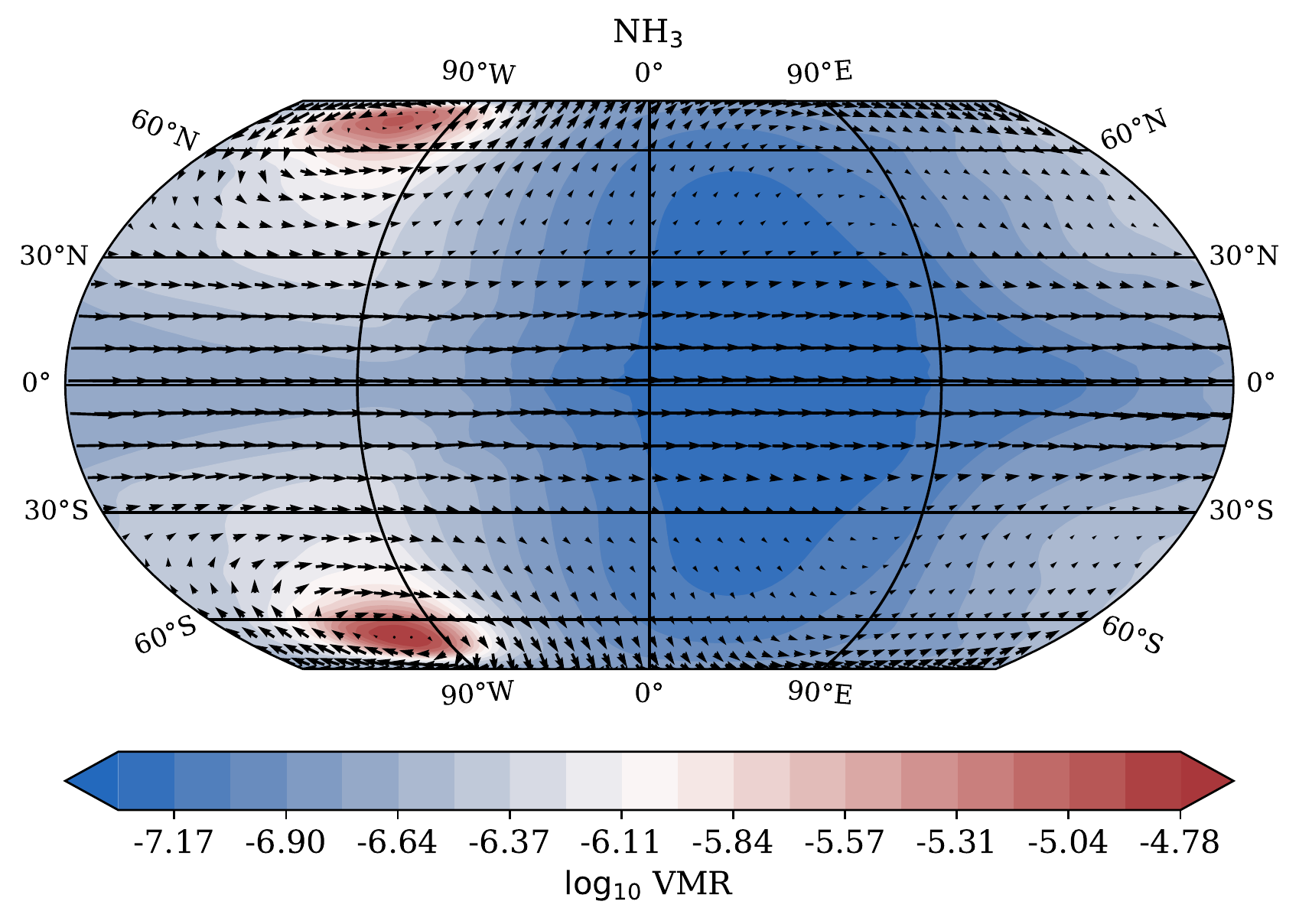} 
   \includegraphics[width=0.45\textwidth]{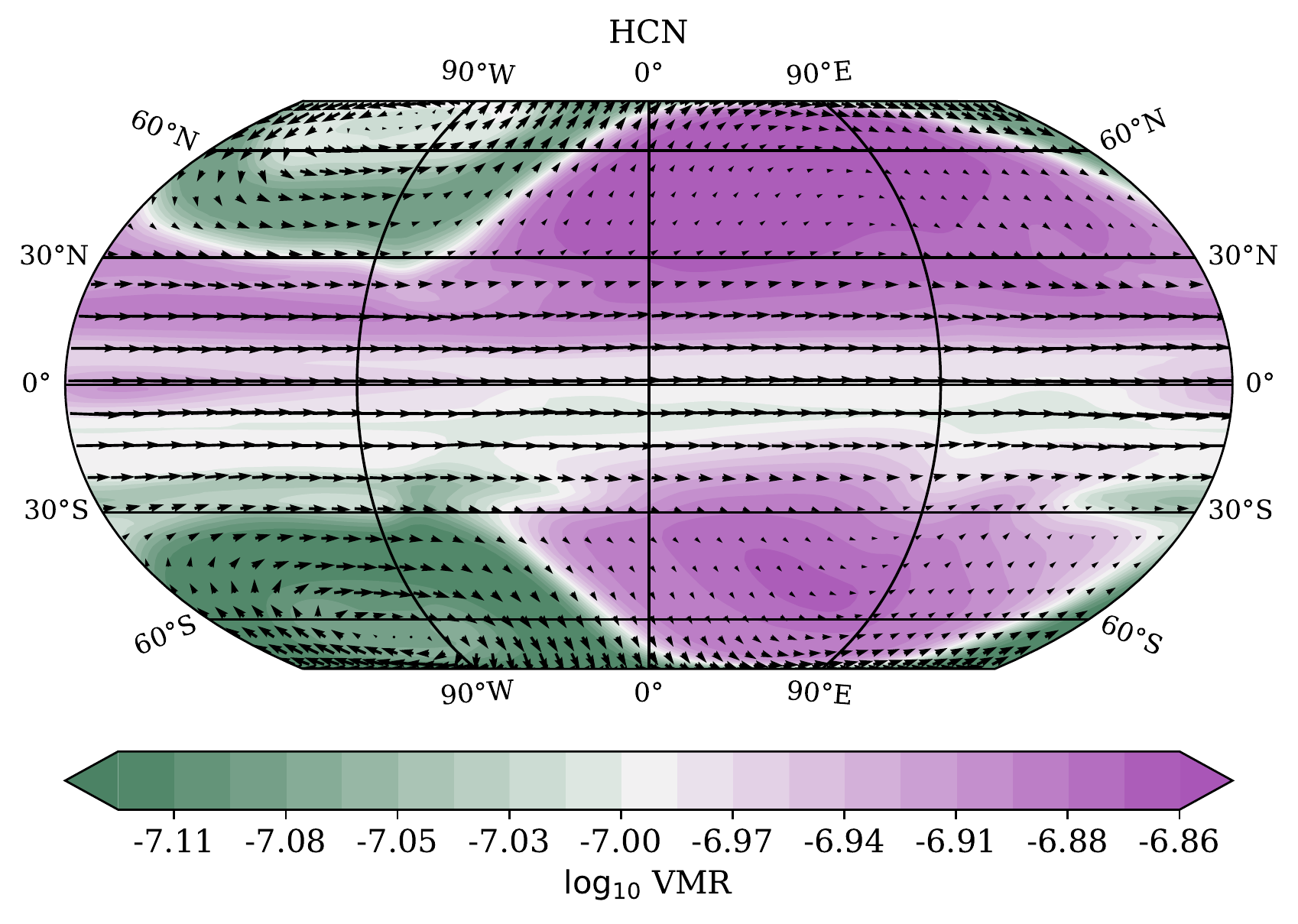}
   \includegraphics[width=0.45\textwidth]{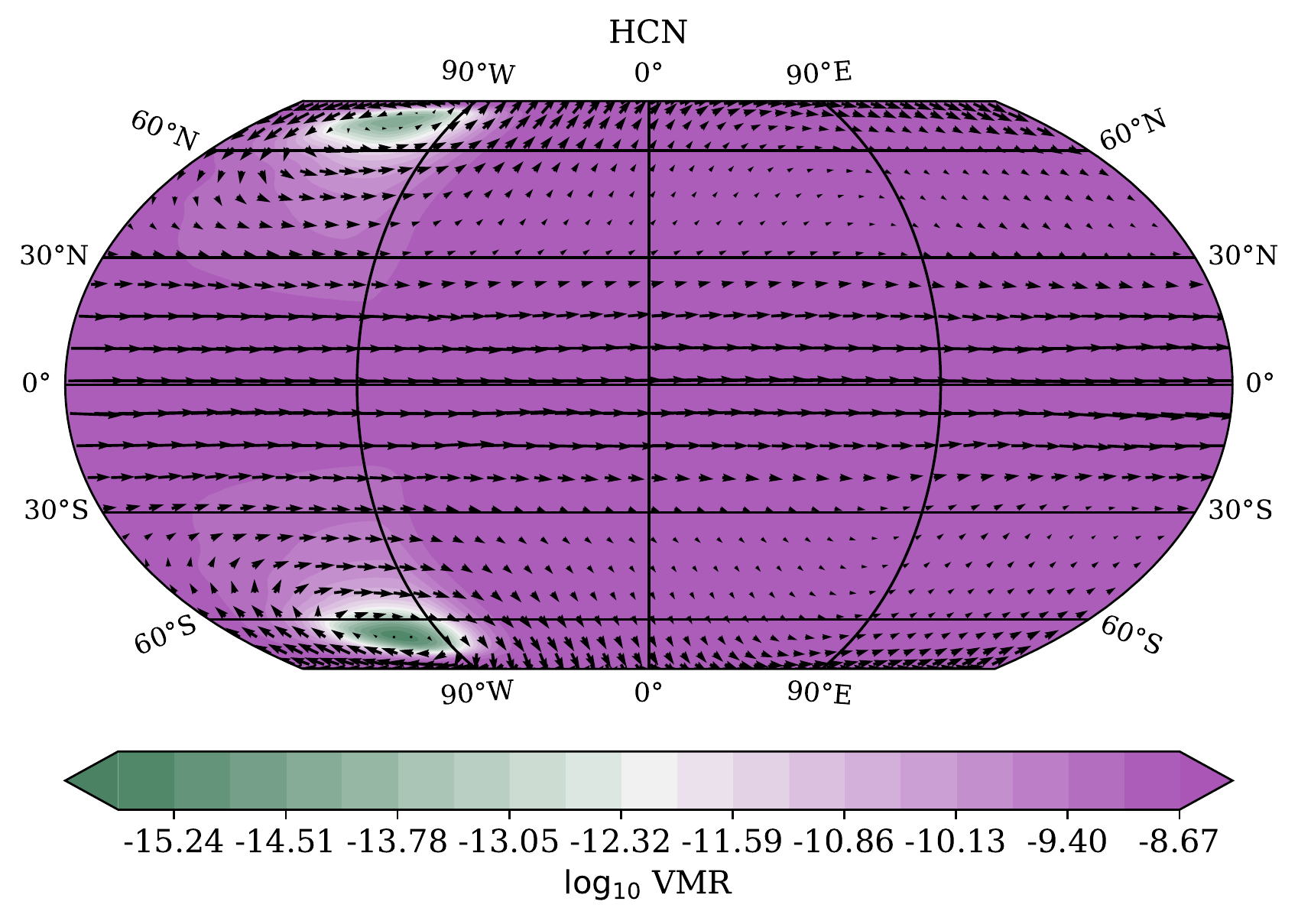}
   \caption{HD 189733b VMR latitude-longitude maps at the 0.1 bar pressure level of the atmosphere, showing the variation of VMR with latitude and longitude of the planet.
   Left: results from the mini-chem coupling.
   Right: results assuming chemical equilibrium.}
   \label{fig:HD189b_map}
\end{figure*}

In Figure \ref{fig:HD189b_temp} we present the vertical temperature-pressure profile and latitude-longitude temperature map at 0.1 bar of the HD 189733b GCM.
Similar to the WASP-39b simulation, this shows a typical HJ circulation pattern but at a slightly higher temperature regime compared to WASP-39b.

In Figure \ref{fig:HD189b_vert} we show the vertical VMR structures as a function of pressure for our four chosen species.
Again, similar conclusions to the WASP-39b simulations can be drawn, where each species has different quench pressures, as well as the quench level being a function of latitude, as seen by the differences between the CE and mini-chem results at the polar regions.
Again, the VMR of each species is highly homogenised at the equatorial regions.
Again, as with the WASP-39b simulation, the deep atmospheric regions, this time below 100 bar, show some non-smooth VMR profiles for CH$_{4}$ and NH$_{3}$.

Figure \ref{fig:HD189b_map} compares the VMR of the four species at the 0.1 bar pressure level between the mini-chem and CE assumptions.
The mini-chem results show a bit of a different distribution of species compared to WASP-39b, with more localised patches of VMR changes with latitude and longitude, corresponding mostly to the dynamical wave driven patterns present in the GCM.
This is very evident in the high latitude Rossby wave patterns on the western and eastern nightside regions.
This pattern of the chemical spatial distribution is similar to that seen in \citet{Drummond2020}.
In this case, the equatorial region is also depleted of CH$_{4}$, CO$_{2}$, NH$_{3}$ and HCN compared to higher latitudes.
This is more in line with spacial distribution found in the \citet{Drummond2020} study.
Again, the VMR maps assuming CE are highly correlated to the temperature field.

\section{Post-processing}
\label{sec:pp}

In this section we post-process our simulations using the gCMCRT 3D Monte Carlo radiative-transfer code \citep{Lee2022b} in transmission and emission. 
Our aim in this section is to not fit the observational data perfectly, but to see the effects of non-equilibrium chemistry on the observational properties of the atmospheres. 

\subsection{WASP-39b}

\begin{figure*} 
   \centering
   \includegraphics[width=0.49\textwidth]{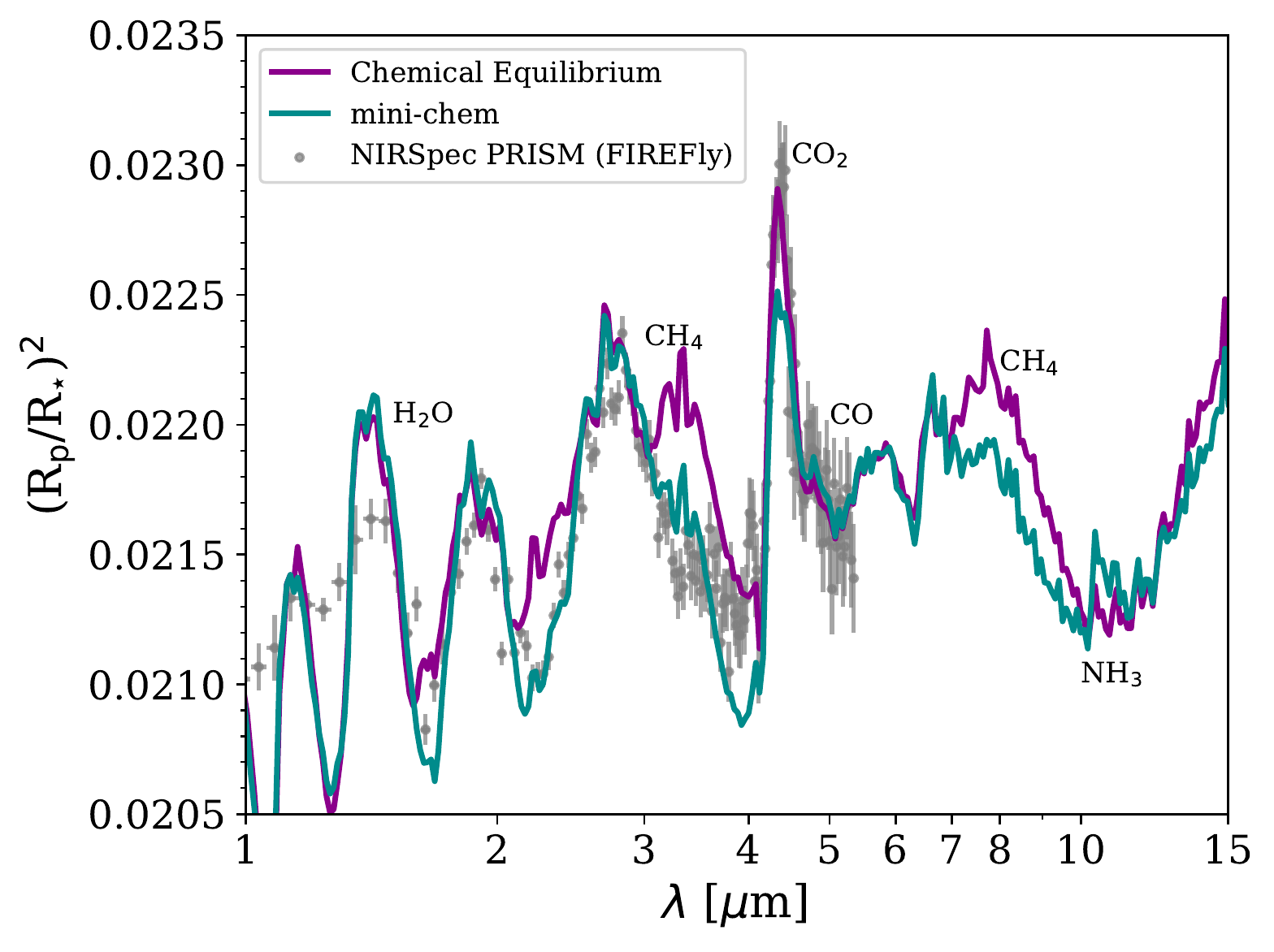}
   \includegraphics[width=0.49\textwidth]{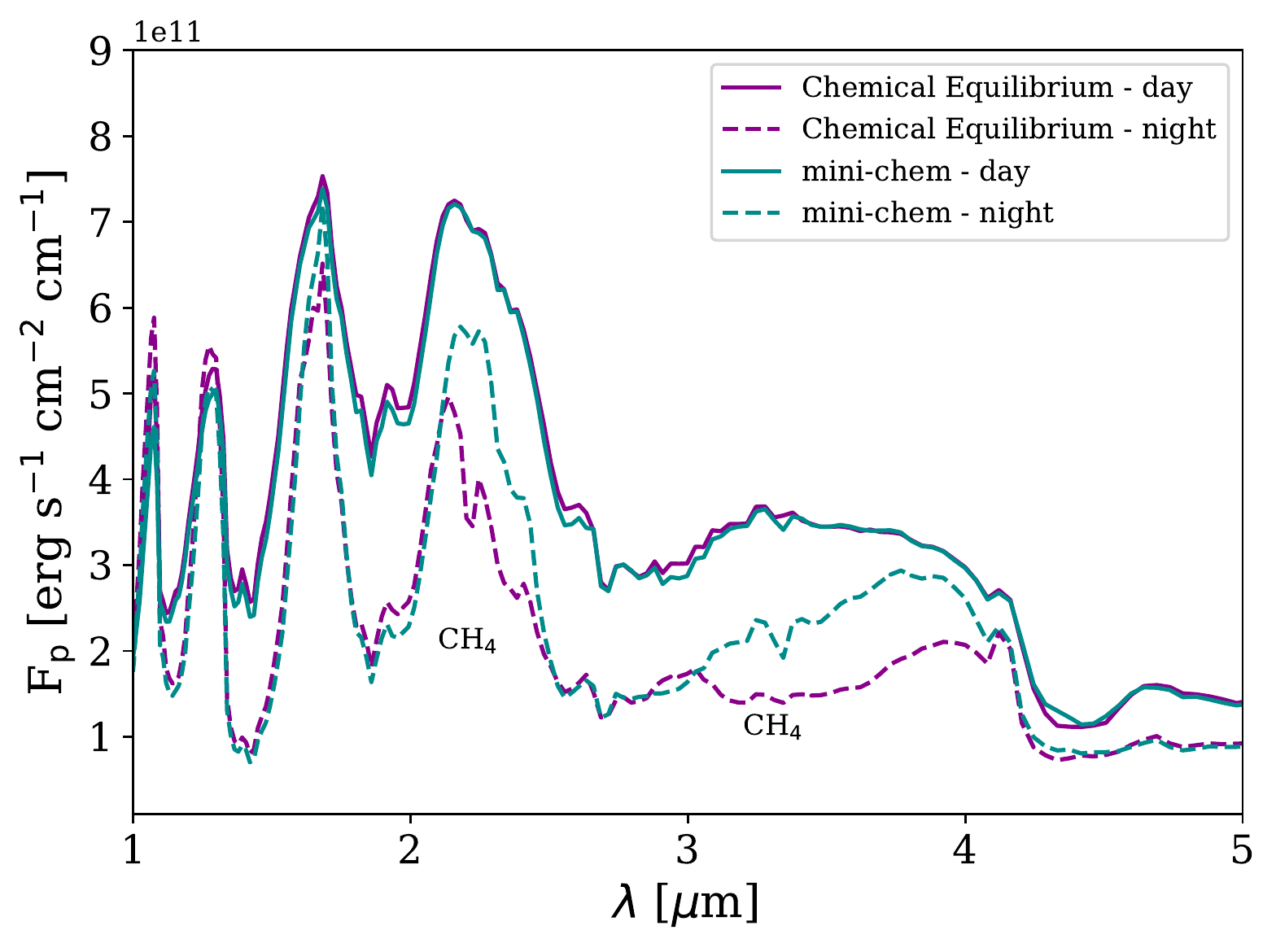}
   \caption{Transmission (left) and emission spectra (right) of the WASP-39b GCM simulation coupled to mini-chem (cyan) and assuming chemical equilibrium (magenta). The NIRSpec PRISM observational data is taken from \citet{JWST2022}.}
   \label{fig:W39b_pp}
\end{figure*}

In Figure \ref{fig:W39b_pp} we present the post-processed transmission and dayside and nightside emission spectra for the WASP-39b GCM model coupled with mini-chem. 
We also produce spectra assuming chemical equilibrium to compare to.
In this case the mini-chem produced transmission spectra contains weaker CH$_{4}$ ($\sim$ 3.3 $\mu$m) and CO$_{2}$ ($\sim$ 4.5 $\mu$m) features and stronger NH$_{3}$ features ($\sim$ 10 $\mu$m). 
We also compare the transit spectra to the JWST NIRSpec PRISM data published in \citet{Rustamkulov2022}. 
Here we see that the reduction in CH$_{4}$ in the kinetic model fits the 3-4 $\mu$m region better but CO$_{2}$ is reduced by too much in the kinetic scheme to fit the CO$_{2}$ feature.
This also has an effect on the emission spectra of the planet on both the dayside and nightside, in particular, the nightside spectrum is greatly affected by the lower amount of CH$_{4}$ around 3.3 $\mu$m, with approximately double the amount of flux able to escape at these wavelengths. 
Interestingly, the dayside emitted flux is more similar between the mini-chem and CE models, suggesting the hottest regions of the exoplanet are well represented by CE.

\subsection{HD 189733b}

\begin{figure*} 
   \centering
   \includegraphics[width=0.49\textwidth]{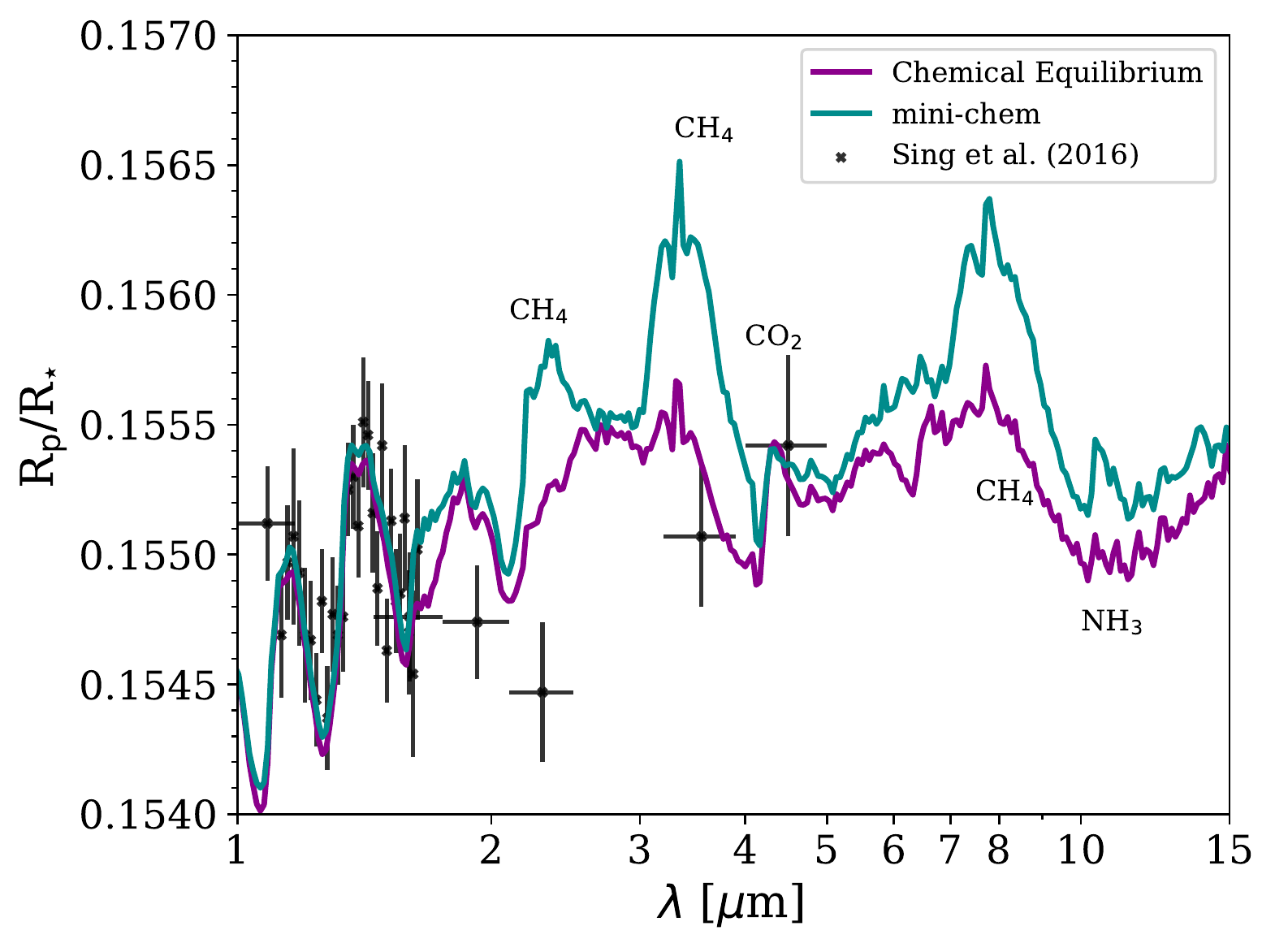}
   \includegraphics[width=0.49\textwidth]{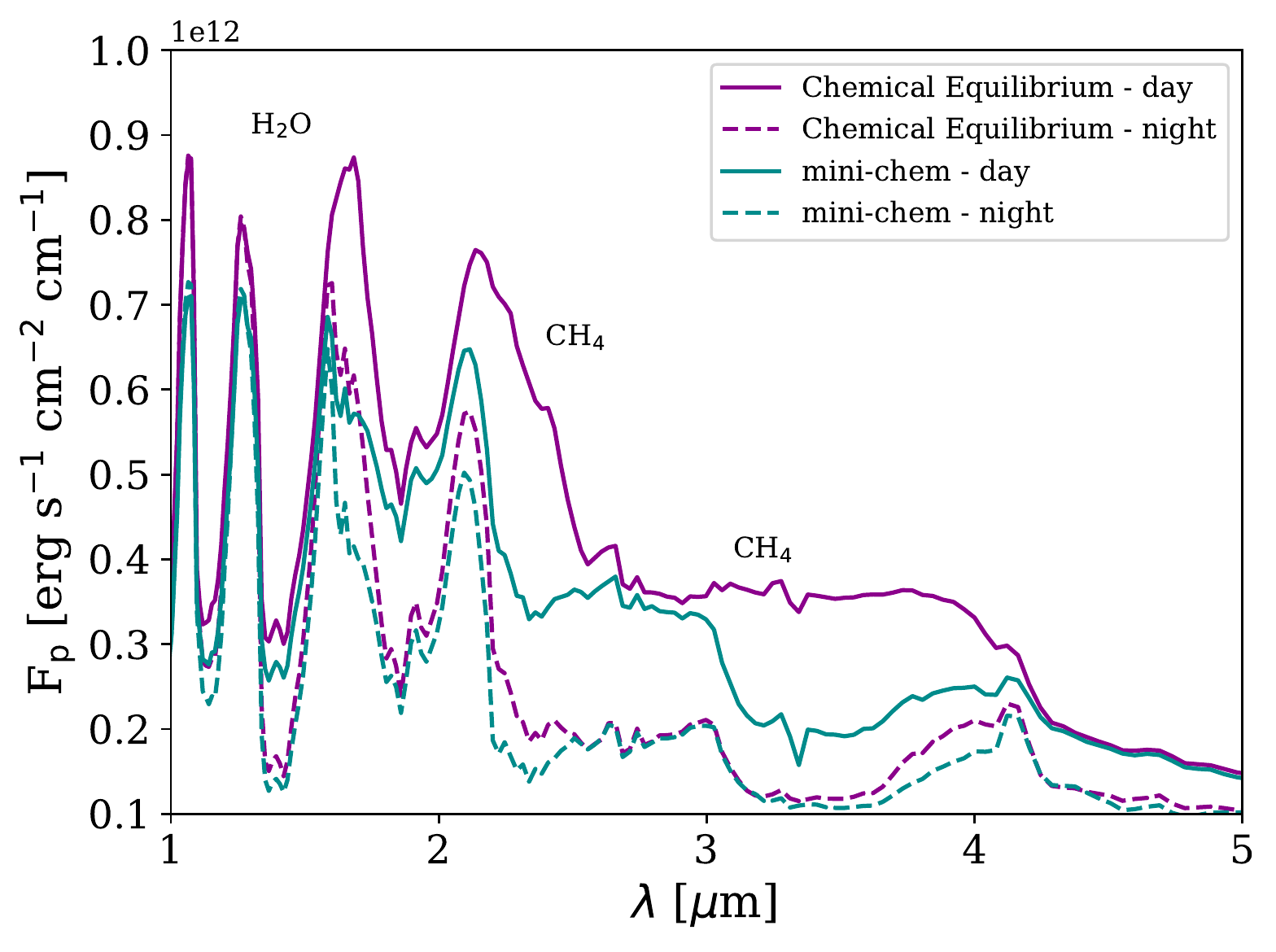}
   \caption{Transmission (left) and emission spectra (right) of the HD 189733b GCM simulation coupled to mini-chem (cyan) and assuming chemical equilibrium (magenta).}
   \label{fig:HD189b_pp}
\end{figure*}

In Figure \ref{fig:HD189b_pp} we present the post-processed transmission and emission spectra for the HD 189733b GCM model coupled with mini-chem. 
We also produce spectra assuming chemical equilibrium as a comparison.
The HD 189733b results show much more dramatic differences to the WASP-39b simulation. 
CH$_{4}$ and NH$_{3}$ spectral signatures are all enhanced in the mini-chem results compared to CE, while the peak of the CO$_{2}$ feature is reduced slightly in the transmission spectra.
We add the current HST and Spitzer data from \citet{Sing2016} to compare the spectra to, suggesting that the large CH$_{4}$ features seen in both the CE and kinetic scheme are not present in the real object.
This may be due to the assumption of Solar M/H used in this study, a higher M/H would reduce the CH$_{4}$ feature size in both cases and also increase the CO$_{2}$ feature, as seen in the WASP-39b case which assumed an M/H = 10x Solar.
Interestingly, in contrast to the WASP-39b models, the dayside emission spectra shows much more difference compared to the nightside spectra.
On the dayside, the flux is generally reduced across the infra-red wavelength regime due to the increased CH$_{4}$ abundance in non-equilibrium. 
Photochemistry may also reduce the dayside CH$_{4}$ abundance substantially, not considered in our study.
 
\section{Comparison to previous studies}
\label{sec:drummond}

The main study that this investigation can be compared to is \citet{Drummond2020} who also coupled a thermochemistry kinetic model to a 3D GCM and performed a HD 189733b model.
Our transmission spectra in Figure \ref{fig:HD189b_pp} show the same change in spectral features between the CE model and the kinetic model as in \citet{Drummond2020}.
This is characterised by a strengthening of the CH$_{4}$ features as well as stronger NH$_{3}$ features in the kinetic model compared to the CE assumption.
We also reproduce the slight decrease of the CO$_{2}$ 4.5 $\mu$m feature in transmission observed in the \citet{Drummond2020} study for the HD 189733b case.
Our emission spectra also agree with the \citet{Drummond2020} study, with the increase in the CH$_{4}$ abundance reducing the emitted flux substantially across the 3-4 $\mu$m range.
The deep atmosphere (p $>$ 10 bar) mixing ratios are also different between each study, primarily due to the use of an initial hot adiabat and higher internal temperature (T$_{\rm int}$ = 382 K here compared to T$_{\rm int}$ = 100 K) for HD 189733b assumed in this study.
Above this deep region, our vertical T-p profiles agree well, despite a simpler picket-fence scheme being used in this study compared to \citet{Drummond2020}'s corr-k model.
However, our simulation produces highly homogenised temperature structures at pressures lower than $\approx$ 0.1 mbar, probably aiding in the homogenisation of the chemical mixing ratios at these low pressures, while \citet{Drummond2020}'s vertical T-p and mixing ratio profiles are truncated near 1 mbar.
Overall, our HD 189733b results agree well with the conclusions of the \citet{Drummond2020} study, the effect of the kinetic model on the emission spectrum is to generally reduce the flux across most of the wavelength range compared to the CE solution. 

\citet{Zamyatina2023} also perform simulations of HD 189733b as part of their exploration of the 3D kinetic chemistry of several planets. 
We find the \citet{Zamyatina2023} results and analysis line up well to our study and the \citet{Drummond2020} study. 
CH$_{4}$ abundance is increased in the 3D case, which produces larger CH$_{4}$ features in the transit spectra, similar to our study, as well as a strong and broad decrease in planetary flux in the CH$_{4}$ band wavelength ranges in emission.

\section{Discussion}
\label{sec:discussion}

Overall, our results point to a dynamically driven chemical composition, where the circulation wave pattern primarily sets the spatial distribution of species in the atmosphere.
In our simulations, each dynamical feature present in the GCM produces a difference in the VMR of certain species with latitude and longitude. 
As well as the equatorial region being enhanced or depleted, dependent on the species and jet structure.
This is in contrast to the CE assumption, which is driven by the local pressure and temperature rather than any dynamical features.

Our results, as well as those presented in \citet{Drummond2020} and \citet{Zamyatina2023}, suggest that non-equilibrium chemistry can affect the conclusions of the physical properties of the atmosphere when drawn from the strength of features in transmission and emission spectra. 
This may lead to inaccurate interpretation of several atmospheric properties without this 3D non-equilibrium consideration.
For example, using 1D RCE models or retrieval modelling to infer atmospheric metallicity and C/O ratios through fitting the strength of H$_{2}$O and CO$_{2}$ or CO features, as well as potential Nitrogen species features, would be affected substantially by the change in global distribution arising from 3D chemical transport effects.
Enhancement of CH$_{4}$ abundances due to dynamical effects would also alter conclusions, and may also provide more CH$_{4}$ to the upper atmosphere for photolysis to dissociate, providing more photochemical products than expected from pure equilibrium effects.

The models presented in this study do not include the effect of radiative feedback from the changing chemical composition due to dynamics.
\citet{Drummond2018b} and \citet{Drummond2020} included this feedback effects for both the chemical relaxation method and the full kinetic scheme respectively.
They found that including radiative-feedback effects can change the atmospheric temperatures on the order of 100 K compared to chemical equilibrium simulations.
A further step from our simulations would be to couple our model to a correlated-k or similar RT scheme to increase the realism of the simulations.
However, the goal of the current project is to showcase the capabilities of the chemistry module itself rather than full self-consistency.

Our mini-chem scheme does not consider the effect of photochemistry on the chemical composition of the atmosphere, shown to be important for setting the overall chemical footprint of the atmosphere \citep[e.g.][]{Hu2021, Baeyens2022}.
Recently, for WASP-39b, signatures of the photochemically produced SO$_{2}$ molecule was seen the JWST NIRSpec PRISM and G395H modes \citep{Rustamkulov2022, Alderson2022}. 
1D photochemical modelling of WASP-39b was able to reproduce the SO$_{2}$ signature \citep{Tsai2023}, suggesting that photochemistry will be a vital consideration for future 3D modelling of exoplanet atmospheres.
Both our chosen planets are ideal for photochemistry studies, with their atmospheric temperatures straddling being between the CO and CH$_{4}$ conversion regimes.
By building on this thermochemistry study, future papers in this series will couple the effects of photochemistry to the net reaction rate scheme to investigate the role in 3D of this important process.
The global distribution of possible haze particle formation due to photochemistry has been recently investigated by \citet{Steinrueck2021}.

Our scheme does not take into account the effects of condensation on the chemical composition of the atmosphere.
For example, the condensation of Silicate Oxides such as Mg$_{2}$SiO$_{4}$ has been shown to reduce the gas phase oxygen abundance \citep[e.g.][]{Helling2008}.
This would affect the C/O ratio and H$_{2}$O abundance in a complex, spatially 3D dependent manner \citep[e.g.][]{Lee2016}, which we foresee will require additional alterations to the current mini-chem scheme to accurately portray.

We note our current GCM models are haze and cloud-free.
Cloud formation is expected to occur in HD 189733b, which has been investigated by a few studies \citep[e.g.][]{Lee2016, Lines2018} and is supported by the current observational data \citep[e.g.][]{Sing2016}.
However, adding microphysical consistent cloud formation and feedback to GCM model has been shown to be a time consuming effort \citep[e.g.][]{Lee2016, Lines2018}, though equilibrium schemes have been proven to be more efficient for running in hot Jupiter GCM models \citep[e.g.][]{Lines2019,Christie2021}.
Clouds can have a large feedback on the temperature structures of hot Jupiters and can affect their observable properties to various degrees dependent on the composition, hemispheric location and vertical depth of the cloud layers \citep[e.g.][]{Lee2017, Christie2021, Parmentier2021,Komacek2022}.  
From the similar T$_{\rm eq}$ of WASP-39b to HD 189733b, it is also likely to form haze and/or cloud particles in its atmosphere \citep{Arfaux2022}.

\section{Summary \& Conclusion}
\label{sec:conclusion}

In this follow up study to \citet{Tsai2022}, we present an open source chemical network, efficient enough for general coupling to hot Jupiter GCMs and other gas giant simulations. 
This network makes use of net reaction tables to reduce the number of species and reactions required to be evolved with the GCM.

We plan to include the effects of photochemistry in future modelling efforts using the net-reaction mini-chem framework.
This will increase the complexity of the model significantly, but will add a much needed chemical processes for warm Neptune planets as well as hot and temperate Jupiters exoplanets.

We have not considered the feedback of the changing VMRs on the radiative-transport in the atmosphere, instead focusing on the chemical scheme and its results alone. 
Our current RT scheme already has this capability which can be examined in detail in future studies. 

Our results suggest that drawing atmospheric bulk properties such as metallicity and C/O ratio from examining the relative strength between absorption features, or fitting using 1D RCE or retrieval modelling, would produce inaccurate results which could rather be attributed to 3D dynamical non-equilibrium chemical effects in the atmosphere.

Overall, this study is further evidence that considering the 3D transport and interaction of gas phase kinetics with the 3D flows is a vital concept for understanding the chemical inventory of exoplanet atmospheres.
Our open-source mini-chem module puts these complexities into the reach of most contemporary GCM modelling groups, without overly taxing computational resources.

\clearpage

\begin{acknowledgements}
E.K.H. Lee is supported by the SNSF Ambizione Fellowship grant (\#193448).
Plots were produced using the community open-source Python packages Matplotlib \citep{Hunter2007}, SciPy \citep{Jones2001}, and AstroPy \citep{Astropy2018}.
The HPC support staff at AOPP, University of Oxford and University of Bern are highly acknowledged.
\end{acknowledgements}

%
\bibliographystyle{aa} 
\bibliography{bib2} 
%

\begin{appendix}

\section{Net reaction contour plots}
\label{sec:mc_contours}

\begin{figure*} 
   \centering
  \includegraphics[width=0.49\textwidth]{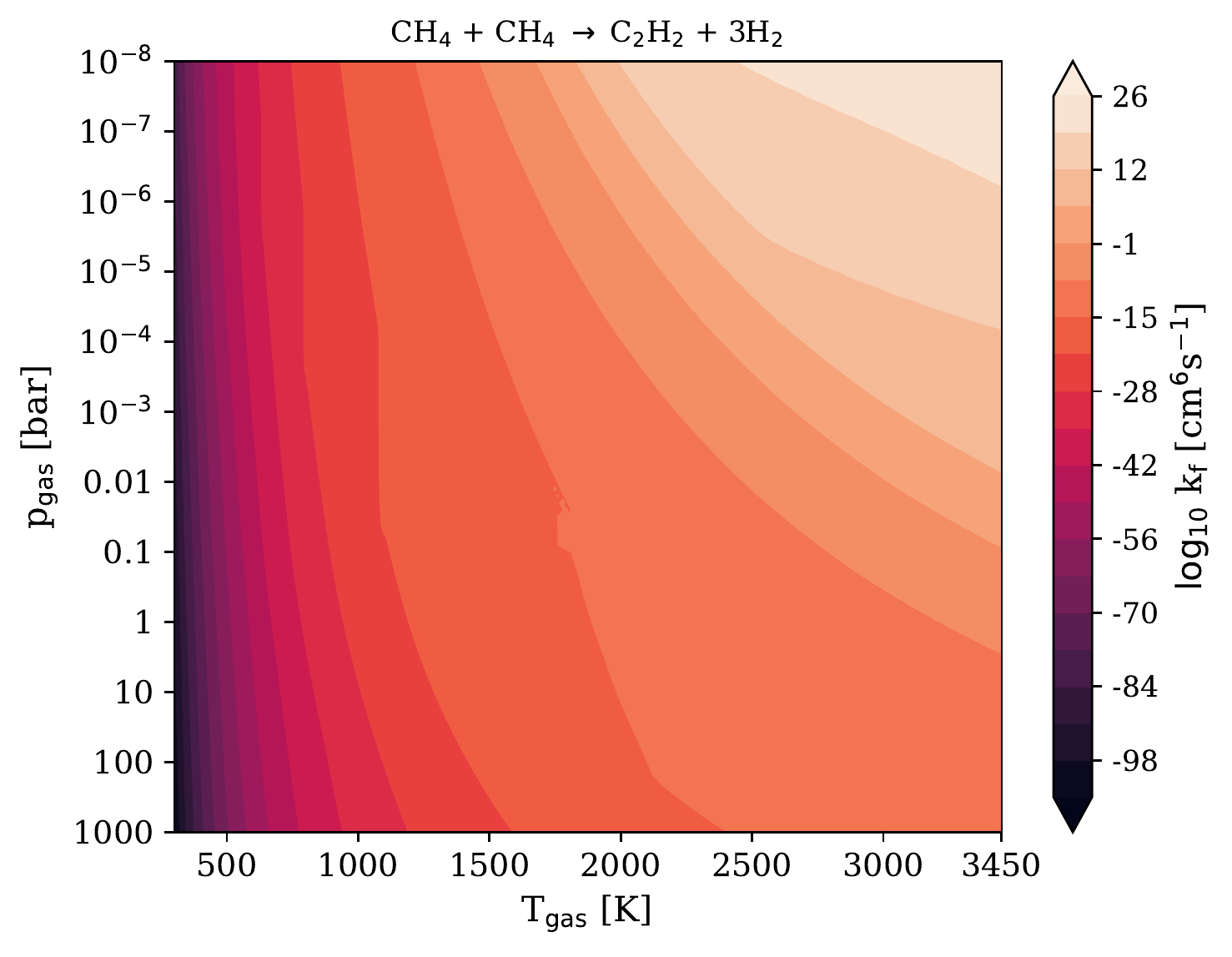}
  \includegraphics[width=0.49\textwidth]{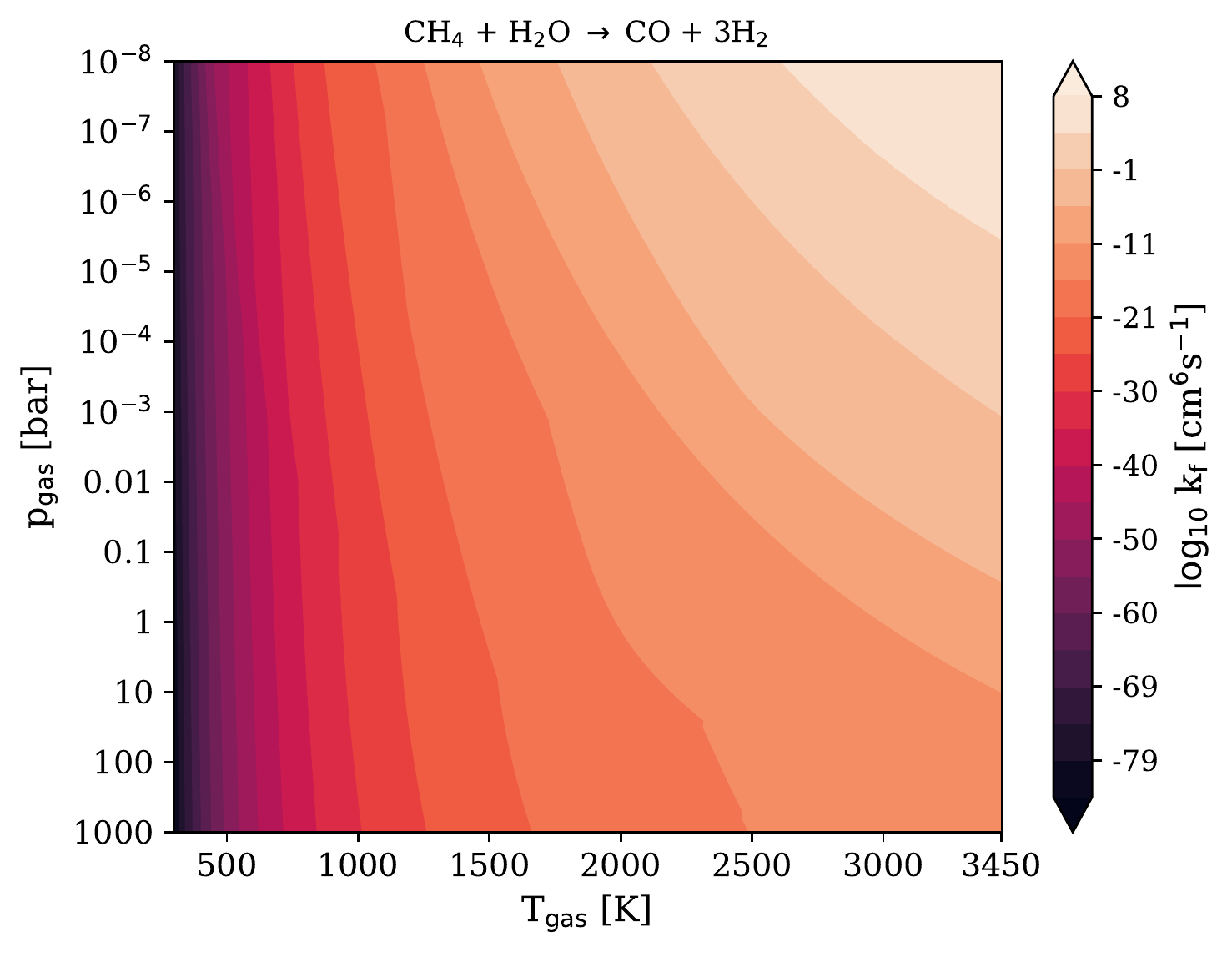}
  \includegraphics[width=0.49\textwidth]{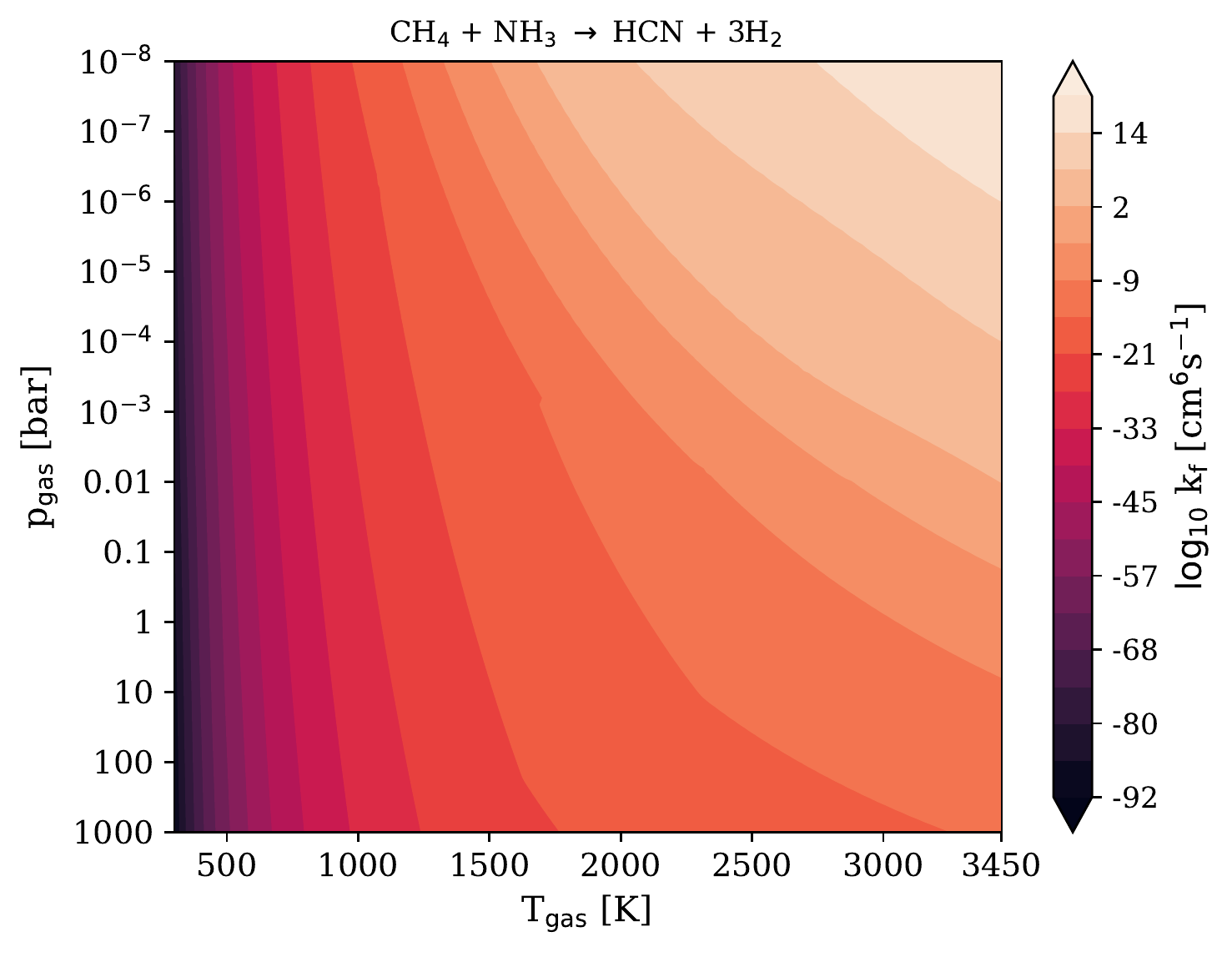}
  \includegraphics[width=0.49\textwidth]{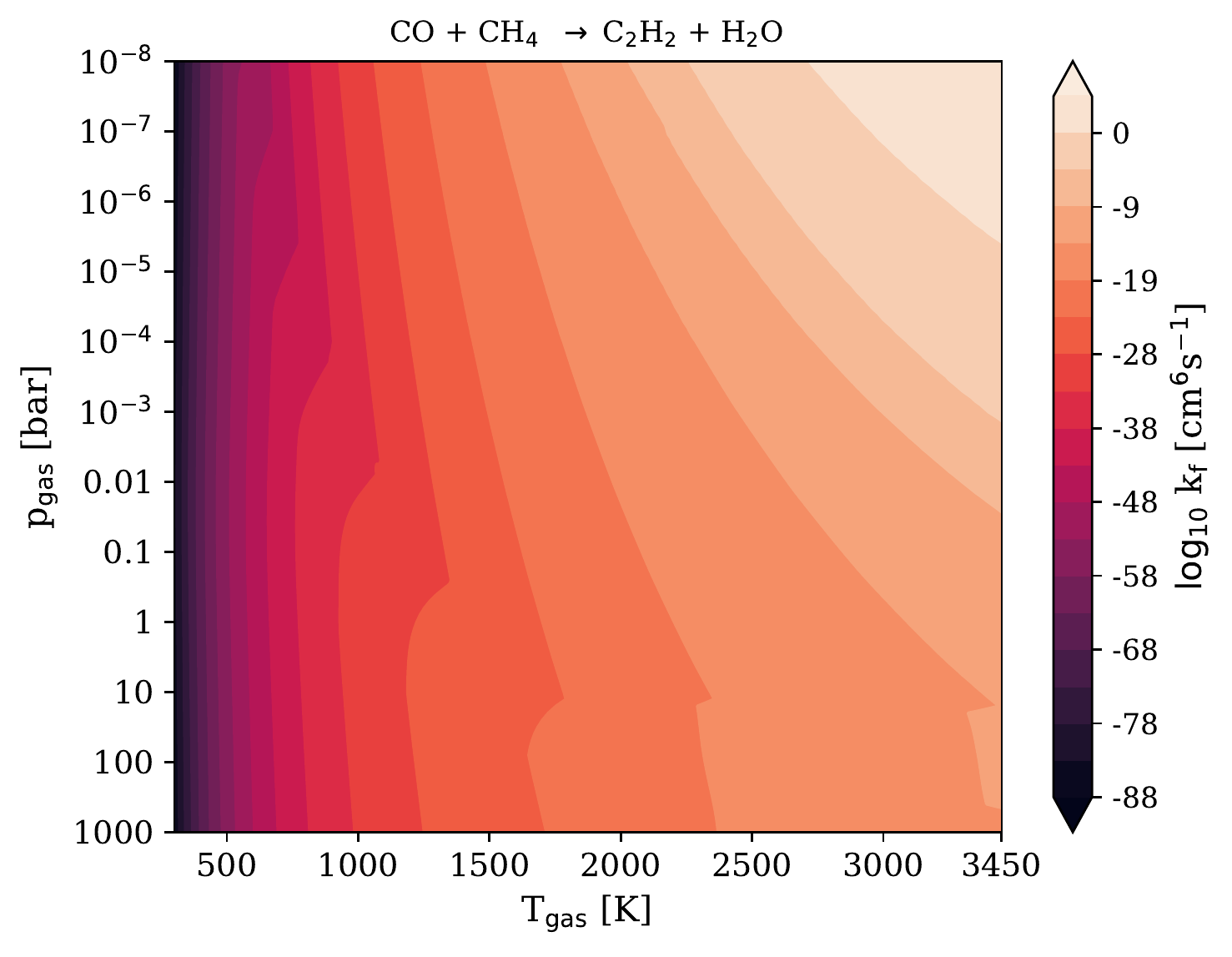}
  \includegraphics[width=0.49\textwidth]{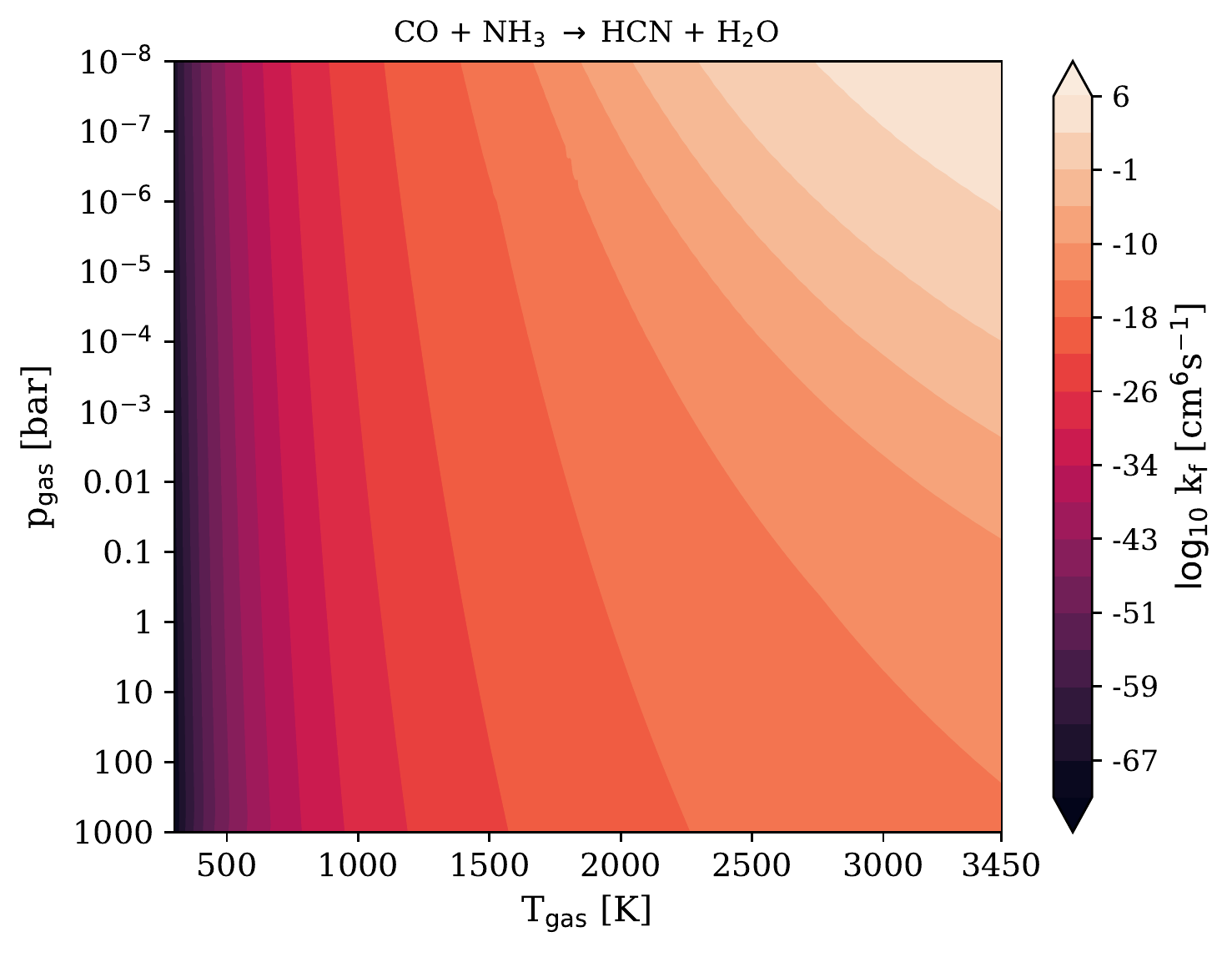}
  \includegraphics[width=0.49\textwidth]{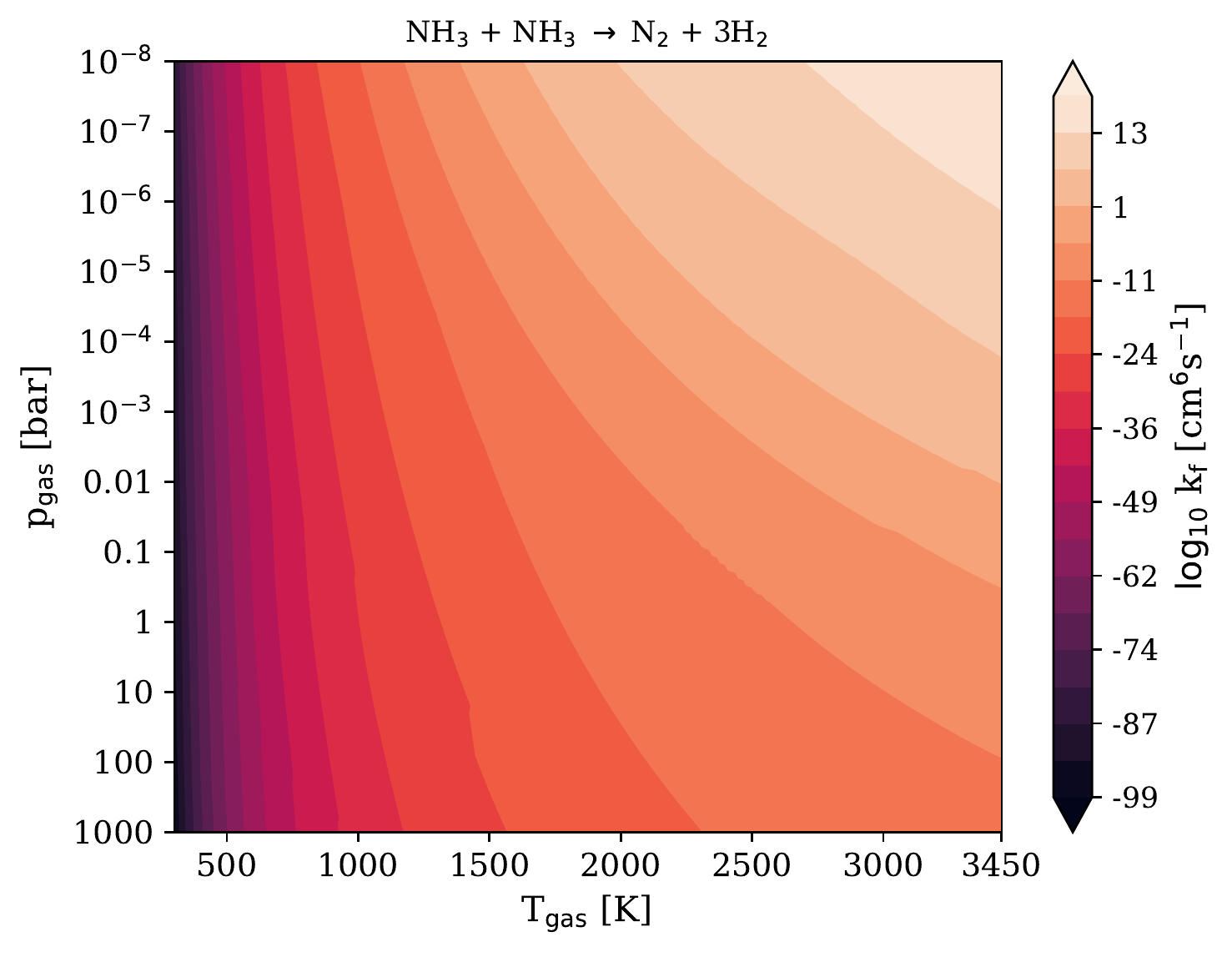}
   \caption{Contour plots of the forward net reaction rate (in $\log_{10}$k$_{\rm f}$) tables as function of gas temperature in Kelvin and total pressure in bar presented in Table \ref{tab:networks}. These plots assume Solar [M/H] values.}
   \label{fig:mc_contours}
\end{figure*}

\end{appendix}

\end{document}